\pdfoutput=1
\documentclass[a4paper,10pt,amsmath,amssymb,nofootinbib,onecolumn,floats,longbibliography,aps,prd]{revtex4-2}

\usepackage[british]{babel}
\usepackage{txfonts}
\usepackage{mathrsfs,bm}
\usepackage{natbib,textcase}
\usepackage{graphicx,color,soul,comment}
\usepackage[pdftex,pdfusetitle]{hyperref}
\hypersetup{colorlinks=true,linkcolor=red,citecolor=blue,filecolor=black,urlcolor=black,
pdfauthor={Edgardo Franzin, Stefano Liberati and Vania Vellucci}}
\usepackage[capitalize]{cleveref}
\usepackage[caption=false]{subfig}
\usepackage{physics}

\usepackage{array,booktabs} %for tables
\usepackage{multirow} %for tables
\newcolumntype{L}{>{$}p{20mm}<{$}} % math-mode version of "l" column type
\makeatletter\g@addto@macro\bfseries{\boldmath}\makeatother%

\allowdisplaybreaks

\def\be#1\ee{\begin{align}#1\end{align}}

\newcommand{\ie}{i.e.}
\newcommand{\eg}{e.g.}

\renewcommand{\dd}{\text{d}}
\newcommand{\e}{\text{e}}

\newcommand{\p}{\partial}
\newcommand{\0}{\nonumber}
\newcommand{\iu}{\mathrm{i}\mkern1mu}

\renewcommand{\geq}{\geqslant}

\renewcommand{\leq}{\leqslant}

\DeclareMathOperator\diag{diag}

\begin{document}

\title{\texorpdfstring{From regular black holes to horizonless objects:\\quasi-normal modes, instabilities and spectroscopy}{From regular black holes to horizonless objects: quasi-normal modes, instabilities and spectroscopy}}

\author{Edgardo Franzin} %\email{efranzin@sissa.it}
\author{Stefano Liberati} %\email{liberati@sissa.it}
\author{Vania Vellucci} %\email{vvellucc@sissa.it}

\affiliation{SISSA, International School for Advanced Studies, via Bonomea 265, 34136 Trieste, Italy}
\affiliation{INFN, Sezione di Trieste, via Valerio 2, 34127 Trieste, Italy}
\affiliation{IFPU, Institute for Fundamental Physics of the Universe, via Beirut 2, 34014 Trieste, Italy}

\begin{abstract}
We study  gravitational and test-field perturbations for the two possible families of spherically symmetric black-hole mimickers that smoothly interpolate between regular black holes and horizonless compact objects accordingly to the value of a regularization parameter.
One family can be described by the Bardeen-like metrics, and the other by the Simpson--Visser metric.
We compute the spectrum of quasi-normal modes (QNMs) of these spacetimes enlightening a common misunderstanding regarding this computation present in the recent literature. 
In both families, we observe long-living modes for values of the regularization parameter corresponding to ultracompact, horizonless configurations. Such modes appear to be associated with the presence of a stable photon sphere and are indicative of potential non-linear instabilities.
In general, the QNM spectra of both families display deviations from the standard spectrum of GR singular BHs.
In order to address the future detectability of such deviations in the gravitational-wave ringdown signal, we perform a preliminary study, finding that third generation ground-based detectors might be sensible to macroscopic values of the regularization parameter.
\end{abstract}

\maketitle

\section{Introduction}

It is widely believed that quantum-gravity effects change the internal structure of black holes (BHs) at some scale $\ell$ and cure the central singularity.
Without specifying the actual theory responsible for these effects, the possible regular spherically symmetric spacetimes can be classified into two families~\cite{Carballo-Rubio:2019fnb} in which the singularity can be either replaced by a global or local minimum radius hypersurface, \ie\ a spacelike wormhole throat hidden inside a trapping horizon, as in the Simpson--Visser (SV) spacetime~\cite{Simpson:2018tsi}, or by an inner horizon shielding a non-singular core~\cite{bardeen_non-singular_1968,hayward_formation_2006,dymnikova_vacuum_1992,Fan:2016hvf}, as the Bardeen-like regular black hole (RBH) models.

For both families, depending on the value of the regularization parameter $\ell$, regular, horizonless configurations can appear~\cite{Carballo-Rubio:2022nuj}.
The SV metric can actually describe a traversable wormhole connecting two symmetric regions of our universe or two universes, while Bardeen-like RBHs can be continuously deformed into horizonless objects.
Remarkably, in addition to the usual unstable photon sphere, these objects may also possess a \emph{stable} photon sphere, whose existence and position depend on the values of the model parameters.

It has been observed that generally linearized perturbations of ultracompact objects with stable photon spheres, as gravastars or constant density stars, decay extremely slowly~\cite{Cardoso:2014sna,Cunha:2017qtt,Cunha:2022gde}.
This strongly suggests that the presence of stable photon spheres can lead to a non-linear instability. This link between long living modes and non-linear instabilities has been recently confirmed by a pseudo-spectrum analysis~\cite{Boyanov:2022ark}.

To investigate whether such instabilities can also be present in the above mentioned horizonless configurations, and to explore the possibility to discriminate the regular from the standard general relativistic BHs through observations, here we study test-field and linear gravitational perturbations in such spacetimes, varying the regularization parameters so to pass smoothly from RBHs to the ultracompact horizonless objects.%
\footnote{An in depth analysis, albeit with a different focus, of the scalar quasi-normal modes of the rotating SV spacetime was carried out in Ref.~\cite{Franzin:2022iai}.}
We stress that regular models are not vacuum solutions of general relativity (GR) and they are often proposed as effective metrics useful for phenomenological analyses.
Nonetheless, in several cases, reverse engineering techniques allow for interpreting these regular models as exact solutions of GR coupled to some suitable matter source ~\cite{Ayon-Beato:2000mjt,Bronnikov:2022ofk,Bronnikov:2021uta,Bronnikov:2022bud,Rodrigues:2023vtm}.
This description, although not unique, makes possible the investigation of \emph{gravitational} perturbations, getting above the study of test-field perturbation on a fixed background.

The continuity between RBHs and horizonless objects of our models enlighten the fact that stable photon spheres are already present in the borderline cases, that is the extremal RBHs and null-throat wormholes, and its position (in Eddington–Finkelstein coordinates) coincides with that of the extremal horizon~\cite{Carballo-Rubio:2022nuj,Carballo-Rubio:2023mvr}. Actually, this is a general feature of any extremal horizon: it coincides with an extreme point of the potential in the equation for null geodesics~\cite{Khoo:2016xqv, Bardeen:1972fi, Pradhan:2010ws}.
 
It has already been argued that extremal BHs should be asymptotically unstable~\cite{Aretakis:2011ha,Aretakis:2012ei}, and it is natural to ask whether the instability associated to the presence of a photon sphere and that associated to extremal horizons, are actually different ways to describe the same phenomenon.
It seems that trapped orbits are indeed present near extremal horizons~\cite{Gralla:2019isj,Ravishankar:2021vip}. While a linear analysis, as the one reported here, is not sufficient to provide a definitive answer to this issue, it can nonetheless enlighten us on the possible relation between the two aforementioned phenomena. In this sense, we investigated if the long-living modes associated with the lightring instability are present also in the extremal case. However, we found that the damping times in the extremal case are orders of magnitude shorter than in the ultracompact cases, suggesting that the photon sphere instability is not triggered or possibly partially suppressed for extremal RBHs.
We argue that this partial suppression could be due to the fact that extremal horizons, being indeed horizons, are not the location of a truly stable orbits but can be considered metastable photon spheres.
Indeed, the presence of a horizon, even if extremal, introduce a source of dissipation, \ie\ the energy that enters the horizon is completely lost.

As a general feature, the QNM spectra for the regular models that we have considered present deviations from the spectrum of a Schwarzschild BH\@.
Assuming that the effect of rotation in more realistic models does not change the picture significantly, we find that for sufficiently large values of the regularization parameter, and for gravitational-wave events with large signal-to-noise ratio, these deviations could be detectable with next generation  detectors~\cite{Punturo:2010zz,LIGOScientific:2016wof,LISA:2017pwj}.

The paper is organized as follows.
In \cref{Models} we describe the two families of regular spacetimes, their main features and field sources.
In \cref{perturb} we illustrate the study of perturbations on these spacetimes, we report the obtained field equations and the methods used to find the corresponding QNMs.
In \cref{results} we show and comment our results, while in \cref{detectability} we discuss how the differences between the obtained spectrum for regular models and the spectrum of singular BHs can be detectable with the next generation of gravitational-waves detectors.
The technical details of the derivation of the perturbative equations are given in the Appendix.

\section{Models\label{Models}}

In this work we consider spherically symmetric and static spacetimes with line element
\be
ds^2 = -\e^{-2 \phi(r)}f(r)\,\dd t^2+\frac{\dd r^2}{f(r)} + r^2\left(\dd\theta^2 + \sin^2\theta\,\dd\varphi^2\right),
\quad f(r) = 1 - \frac{2m(r)}{r}\,.
\label{metric}
\ee

Most of the Bardeen-inspired models are described by the line element \eqref{metric} with $\phi(r)=0$ and some given mass function $m(r)$ which contains a regularization parameter $\ell$.
Besides Bardeen's original proposal, a non-comprehensive list of widely explored models includes Hayward~\cite{hayward_formation_2006},
Dymnikova~\cite{dymnikova_vacuum_1992}, Fan--Wang~\cite{Fan:2016hvf}, and many others.
A notable exception is the SV spacetime which is obtained from the Schwarzschild line element with the substitution $r\to\sqrt{r^2+\ell^2}$. After a change of coordinates, even its line element can be written in the form \eqref{metric}.
As examples of the two possible families of regular geometries, in this paper we consider the Bardeen and SV models,
\begin{alignat}{3}
&\text{Bardeen:} & \phi(r) &= 0\,, & m(r) &= M \frac{r^3}{(r^2 + \ell^2)^{3/2}}\,,\\
&\text{Simpson--Visser:}\quad & \phi(r) &= \frac{1}{2}\ln\left(1 - \frac{\ell^2}{r^2}\right)\,,\quad & m(r) &= M\left(1 - \frac{\ell^2}{r^2}\right) + \frac{\ell^2}{2r}\,.
\end{alignat}

Remarkably, it can be shown that these two families of solutions pretty much cover all the possible regularized, spherically symmetric, static, BH spacetimes~\cite{Carballo-Rubio:2019fnb}.~\footnote{In Ref.~\cite{Carballo-Rubio:2019fnb} where also found spacetimes with an asymptotic regularization of the singular behaviour, here we stick to the more physically realistic ones in which the regularizing effects due to quantum gravity are supposed to act on a finite region of spacetime.} Furthermore, depending on the value of $\ell$, they can also describe ultra-compact, horizonless, objects~\cite{Carballo-Rubio:2022nuj} which will also be considered in this study.

\subsection{Horizons and photon spheres}

\begin{figure*}[ht]
\centering
\includegraphics[width=0.42\textwidth]{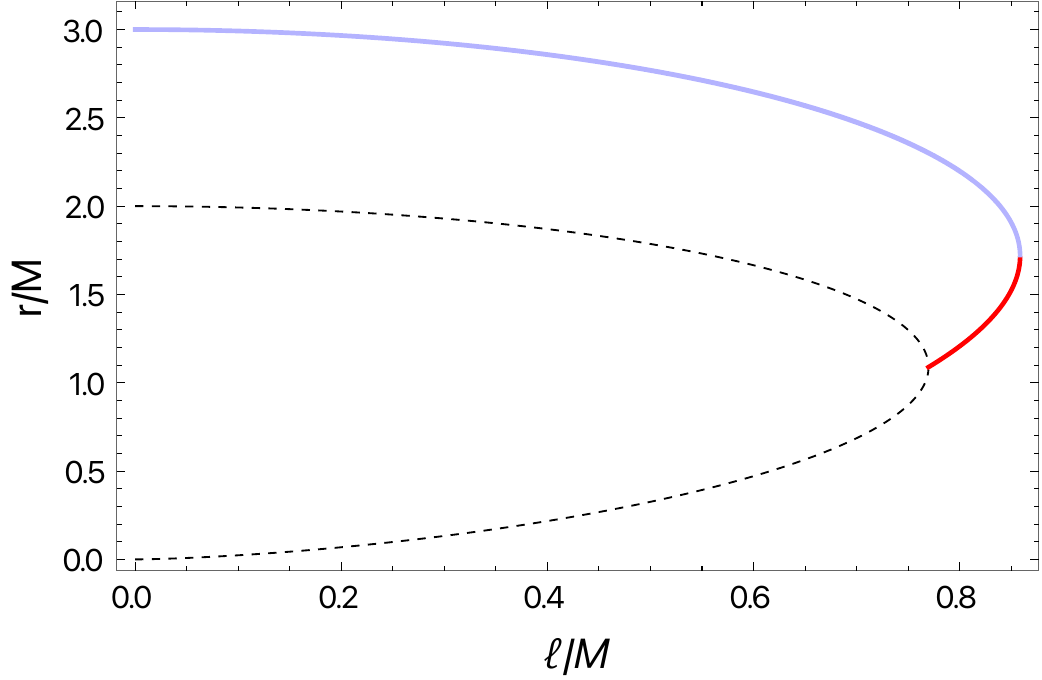}
\qquad
\includegraphics[width=0.42\textwidth]{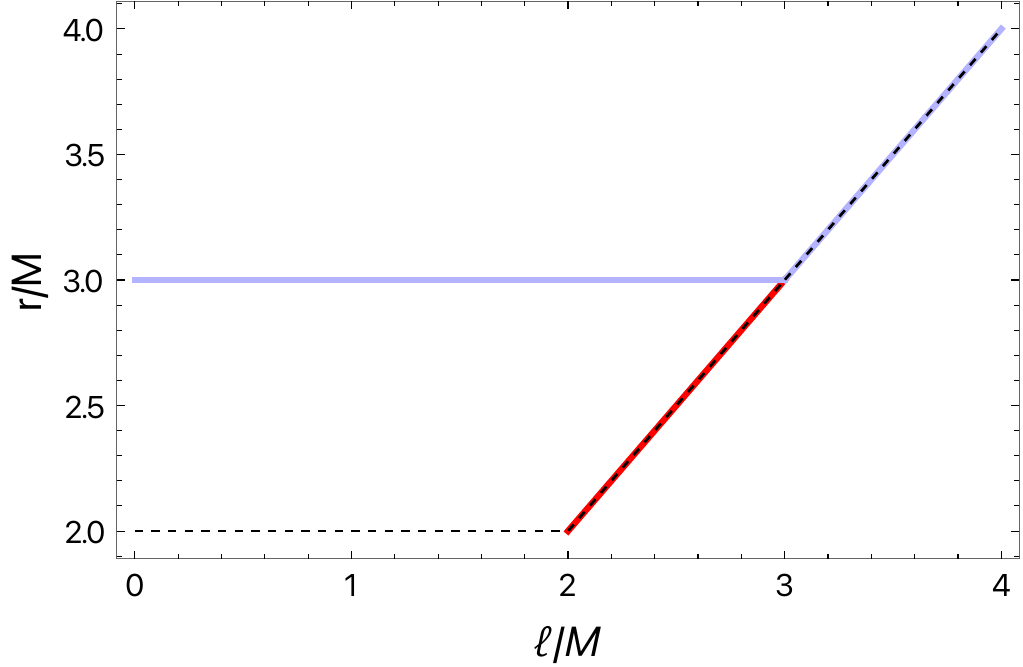}
\caption{Radii of the photon spheres (solid red lines for the inner stable one and solid purple line the outer unstable one) and horizons (dashed black line) for the Bardeen (left panel) and SV (right panel) metric.
For the Bardeen metric the two horizons merge for $\ell =4M/3 \sqrt{3}$ giving way to a stable photon sphere inside the usual unstable one. For $\ell = \frac{48 M}{25 \sqrt{5}}$ the two photon spheres finally merge leaving a simple compact object. For the SV metric the horizon becomes a wormhole throat for $\ell =2M$  over which a stable photon sphere resides. For $\ell=3M$ the two photon spheres merge and the wormhole throat becomes an unstable photon sphere.}
\label{fig:structure}
\end{figure*}

A first relevant observation for our analysis is that the Bardeen-like and SV models have very different features when interpolating from RBHs to horizonless objects. However, in both cases, there exist two special values of the regularization parameter $\ell$, say $\ell_\text{ext}$ and $\ell_\text{light}$ with $\ell_\text{ext}<\ell_\text{light}$, which determine the existence and position of horizons and photon spheres.
This is visually illustrated by comparing \cref{fig:structure}, where the horizon and photon-sphere structure of the spacetimes is represented according to the value of $\ell$.

On the one hand, for $\ell<\ell_\text{ext}$ a Bardeen-like line element describes a RBH with two horizons and one unstable photon sphere;
for $\ell=\ell_\text{ext}$ the spacetime becomes an extremal RBH, in which the two horizons and the unstable photon sphere coincide;
for $\ell_\text{ext}<\ell<\ell_\text{light}$ the horizon disappears, the spacetime describes an ultracompact object with two photon spheres whose distance decreases with increasing values of $\ell$, and one of them is stable.
Finally, for $\ell>\ell_\text{light}$ the two photon spheres disappear and nor stable or unstable null circular orbits are anymore possible around the object.
In particular, for the Bardeen model those special values read $\ell_\text{ext}/M = \frac{4}{3\sqrt{3}}$ and $\ell_\text{light}/M=  \frac{48}{25 \sqrt{5}}$. 

On the other hand, the SV metric for $\ell<2M$  describes a RBH geometry with a single horizon shielding a one-way spacelike throat, surrounded by an unstable photon sphere;
for $\ell=2M$ the spacetime represents a one-way wormhole with an extremal null throat and two photon spheres, one of which is stable and located at the throat;
for $2M<\ell<3M$ the wormhole becomes traversable both ways, the throat at $r=0$ is timelike and there are two accessible photon spheres; for $\ell>3M$ the spacetime has only one unstable photon sphere located at the throat $r=0$.

\subsection{Field sources\label{s:sources}}
As said before, the above introduced static solutions, can be considered the outcome of a transient regularization of the gravitational collapse due to quantum gravity. The implicitly assumption is that such non-classical regime gives way, at late times, to a stationary configuration that should be a solution a some gravitational theory: a low energy, effective field theory limit of quantum gravity, whatever this might be. As our solutions mimic GR ones better and better as one gets away from the objects cores, so we do expect that any such effective field theory of gravity should be encoding deviations from GR in strong gravity regimes. Also, it is well known that such theories can often be recast as GR with non-trivial, and sometimes exotic, matter sources. It is hence reasonable to explore the interpretation of our geometries as solutions of GR and check their associated matter content as this is a crucial step for considering their behaviour under perturbations.

Within GR, the effective stress-energy tensor associated with the line element \eqref{metric} is given by its Einstein tensor, \ie, $T^\mu{}_\nu = G^\mu{}_\nu/8\pi$. Then, for any given RBH model, one might question {\em a posteriori} the existence of some matter distribution yielding the same stress-energy tensor.

Notice that the Einstein tensor computed from \cref{metric} has three independent components,  meaning that the matter source cannot be uniquely a scalar field (for which $T^t{}_t=T^\theta{}_\theta$), nor an electromagnetic field (for which $T^t{}_t=T^r{}_r$).

Nonetheless, when $\phi(r)=0$, $G^t{}_t=G^r{}_r$ and Bardeen-like RBHs are often interpreted as solutions of GR coupled to some non-linear electrodynamics with action~\cite{Ayon-Beato:2000mjt,Bronnikov:2022ofk}
\be
\mathcal{S} =\int \dd^4 x\,\sqrt{-g}\left(\frac{1}{16 \pi} R-\frac{1}{4 \pi} \mathcal{L}(F)\right),
\label{action}
\ee
where the electromagnetic Lagrangian is a non-linear function of the electromagnetic field strength $F=\frac{1}{4}F_{\mu\nu}F^{\mu\nu}$, with $F_{\mu\nu} = 2\nabla_{[\mu}A_{\nu]}$ being $A_\mu$ the electromagnetic potential. The Maxwell field is frequently assumed purely magnetic and its magnetic charge coincides with the regularization parameter, which implies that the only non-vanishing component of the Maxwell field is $F_{\theta\varphi}=\ell\sin\theta$ (alternatively, the only non-vanishing component of the potential is $A_\varphi = \ell\cos\theta$) and $F=\ell^2/2 r^4$.

The modified Maxwell field equation
\be
\nabla_\mu\left(\mathcal{L}_F F^{\alpha \mu}\right)=0\,,\label{EqMotMax}
\ee
being $\mathcal{L}_F\equiv\p\mathcal{L}/\p F$, is trivially satisfied, while the gravitational equations
\be
G_{\mu\nu} = 2\left(\mathcal{L}_F F_{\mu}{}^\lambda F_{\nu\lambda} - g_{\mu\nu} \mathcal{L}\right),\label{EqMotGrav}
\ee
imply that the electromagnetic Lagrangian is given in term of the metric functions of the spacetime as in \cref{metric} (with $\phi=0$)
\be
\mathcal{L}(F) = \frac{m'}{r^2},
\ee
where $r=r(F)$.

In particular, for the model considered in this paper
\be
\mathcal{L}_\text{Bardeen} = \frac{3 M}{\ell^3} \left(\frac{\sqrt{2 \ell^2 F}}{1 + \sqrt{2 \ell^2 F}}\right)^{5/2}\,.
\ee

On the other hand, when $\phi\neq0$, to model the source it is necessary to introduce other matter fields.
In particular, the SV spacetime could be sourced by a combination of non-linear electrodynamics and a self-interacting scalar field~\cite{Bronnikov:2021uta,Rodrigues:2023vtm}.
\be
\mathcal{S} =\int \dd^4 x\,\sqrt{-g}\left(\frac{1}{16 \pi} R-\frac{1}{4 \pi} \mathcal{L}(F) - \frac{\varepsilon}{2}(\p\Phi)^2 - V(\Phi)\right),
\label{actionSV}
\ee
where $\varepsilon= \pm1$, and the positive (negative) sign corresponds to a canonical (phantom) scalar field with positive (negative) kinetic energy.

Even in this case, we assume the Maxwell field to be purely magnetic with its magnetic charge equal to the regularization parameter, so that the modified Maxwell equation is trivially satisfied.
The computation of the gravitational field equations
\be
G_{\mu\nu} = 2\left(\mathcal{L}_F F_{\mu}{}^\lambda F_{\nu\lambda} - g_{\mu\nu} \mathcal{L}\right)
+ 8\pi\left[\varepsilon\p_\mu\Phi\p_\nu\Phi - g_{\mu\nu}\left(\frac{\varepsilon}{2}(\p\Phi)^2 + V(\Phi)\right)\right],\label{EqMotGravSV}
\ee
reveals that the scalar field is phantom and satisfies
\be
\Phi'^2 = \frac{\phi'}{4\pi r}\,,
\ee
the derivative of the electromagnetic Lagrangian reads
\be
\mathcal{L}_F = \frac{r^2 \left[r^2 f''-3 r^2 f' \phi' - 2 f \left(r^2 \phi'' - r^2 \phi'^2 + r \phi' + 1\right) + 2\right]}{4 \ell^2}\,,
\ee
which, once integrated, can be substituted in the expression for the scalar potential
\be
V[\Phi(r)] = \frac{1-r f' + f \left(r \phi'-1\right)-2 r^2 \mathcal{L}}{8\pi r^2}\,.
\ee
Finally, the Klein--Gordon equation is a consequence of the Einstein equations.

In particular, using the metric functions for the SV spacetime we get
\be
\mathcal{L}_\text{SV} = \frac{6M}{5} \left(\frac{2 F^5}{\ell^2}\right)^{1/4}\,,\quad
\Phi_\text{SV} = \frac{1}{\sqrt{4\pi}} \arccot \frac{\sqrt{r^2-\ell^2}}{\ell}\,,\quad
V_\text{SV} = \frac{M \sin^5\left(\sqrt{4\pi}\,\Phi \right)}{10\pi \ell^3}\,.
\ee

Notice that we have used a different convention with respect to Ref.~\cite{Bronnikov:2021uta}; in particular we have chosen the scalar field to vanish at spatial infinity.

\section{Study of perturbations\label{perturb}}

Assuming the gravito-scalar-magnetic interpretation given in \cref{s:sources}, we can study the full effect of linear perturbations expanding the metric and the matter fields around their background values.
According to their parity symmetry, even or odd, the metric and matter perturbations can be decomposed respectively in polar and axial contributions.
However since the background metric and the background scalar field are even, while the background magnetic field is odd, axial electromagnetic perturbations and polar scalar perturbations are coupled to polar gravitational perturbations, while polar electromagnetic perturbations are coupled solely to axial gravitational perturbations (being impossible to have axial scalar perturbations).

If this parity coupling is not taken into account, as it commonly happened in several recent investigations~\cite{Ulhoa:2013fca,Dey:2018cws,Toshmatov:2018tyo,Toshmatov:2018ell,Toshmatov:2019gxg,Zhao:2023jiz,Wu:2018xza}, one obtains an incompatible system of equations for the perturbation functions, which admits only a trivial solution.
This has been quite systematically overlooked in the previously mentioned literature, tacitly assuming that one of the equations can be obtained from the other two.
To understand the fine details, the interested reader can follow the full derivation of the perturbative equations in \cref{app:perturbations} and in particular the comment in \cref{footnote:wrong}.
On the other hand, other authors have analyzed linear perturbations carefully, but specialized to non-linear electrodynamics without scalar fields or viceversa~\cite{Bronnikov:2012ch,Moreno:2002gg,
Chaverra:2016ttw,Nomura:2020tpc,Nomura:2021efi,Meng:2022oxg}.
Our perturbative analysis extends these results to a generic spacetime described by the line element~\eqref{metric}, interpreted as an exact solution of GR coupled to non-linear electrodynamics and scalar fields.

\subsection{Full perturbative analysis}

For each parity sector, gravitational, scalar and electromagnetic harmonic perturbations satisfy a system of coupled non-homogeneous wave equations, which schematically read
\be
\frac{\dd^2\mathcal{I}}{\dd r_*^2} + \left(\omega^2- V_\mathcal{I} \right)\mathcal{I}
+ \sum_{\mathcal{J\neq I}} c_\mathcal{I,J}\,\mathcal{J} &= 0\,,\label{eqpert_gen}
\ee
where $r_*$ is the tortoise coordinate defined as $\dd r_*/\dd r \equiv \e^{\phi}/f$, for $\mathcal{I,J=\{A,E\}}$ in the sector in which axial gravitational perturbations are coupled to polar electromagnetic perturbations, and $\mathcal{I,J=\{P,B,S\}}$ in the sector in which polar gravitational, axial electromagnetic and polar scalar perturbations are coupled.
The variables $\{\mathcal{A,P,B,E,S}\}$ are given combinations of the metric, the electromagnetic potential, and the scalar field perturbation functions and their derivatives.
The potentials $V_\mathcal{I}$ and the coefficients $c_\mathcal{I,J}$ are given functions of the background metric and fields, and also depend on the harmonic number $l$ associated to the spherical-harmonics expansion.

For the sake of conciseness, we shall only summarise, in~\cref{results}, the outcome of such full perturbative analysis, which is instead explicitly carried on in \cref{app:perturbations}. The latter turns out to be quite involved and dependent on the details of the matter distribution so, as a complementary analysis, we also present in what follows a test-field perturbations analysis which, albeit less accurate, has the merit to avoid assumptions on the matter distribution supporting the geometry. In~\cref{results} we shall see that, reassuringly, the outcomes between the two kinds of analysis turn out to be qualitatively in agreement.

\subsection{Test-field perturbations\label{s:testfpert}}

Dealing with metrics for which a specific distribution of matter is not specified, test-field perturbations represent a simple but informative proxy.
Often, the first step is to consider scalar field perturbations on top of these spacetimes.
For spherically symmetric spacetimes, it is possible to extend the analysis to other spin-$s$ fields, to include \emph{axial} gravitational spin-2 perturbations~\cite{Medved:2003pr}.

The crucial point made in such analyses is that standard matter fields --- such as canonical scalars and Maxwellian electric fields --- couple with the \emph{polar} gravitational perturbations, while in the axial sector the source stress-energy tensor is left unperturbed. If  true this would imply that the axial gravitational sector could already capture some features of the QNM spectrum. However, this is not the case for a purely magnetic source, and one should be then careful in drawing conclusions.

Within this context, the perturbative equation for scalar, electromagnetic and gravitational axial perturbations for the spacetime described by \cref{metric} reads~\cite{Karlovini:2001fc,Medved:2003pr}
\be
\frac{\dd^2 \psi_s}{\dd r_*^2} + \left(\omega^2 - V_s\right)\psi_s = 0\,,
\label{HarmEq2}
\ee
where $\psi_s$ is related to the spin-$s$ perturbed functions, the tortoise coordinate $r_*$ is still defined as $\dd r_*/\dd r \equiv \e^{\phi}/f$, and the potential depends on the spin-weight of the perturbation and the metric functions
\be
V_s = f\,\e^{-2\phi} \left[\frac{l(l+1)}{r^2}+\frac{2 \left(1-s^2\right) m}{r^3}-(1-s)
\left(\frac{2 m'}{r^2}+\frac{f \phi'}{r}\right)\right].
\ee

It is interesting to note the formal similarity between equations \cref{eqpert_gen,HarmEq2} modulo the last term on the r.h.s.\ of \cref{eqpert_gen}.

\subsection{Computation of the quasi-normal modes}

We now want to solve \cref{eqpert_gen,HarmEq2} for $\omega$, to compute the quasi-normal mode (QNMs) frequencies, \ie\  the late-time response of the compact object to an initial perturbation that is localized in space.
After providing suitable boundary conditions, that depend on the physical process and on the compact object properties, we use standard and matrix-valued direct integration techniques~\cite{Chandrasekhar:1975zza,Pani:2013pma} for the test-field case and for the full gravitational case, respectively.

For the RBH cases, the two boundaries from which we integrate are spatial infinity, where we impose the solution to be a purely outgoing wave, and the horizon, where we impose the solution to be a purely ingoing wave.

For the horizonless cases, we still impose the solution to be purely outgoing at spatial infinity, but for the other boundary condition we make a different choice for the two families.
The Bardeen-like metrics with $\ell > \ell_\text{ext}$ describe ultracompact stars thus we impose regularity conditions at the origin.
Note that in this way we are assuming that the test field perturbations can travel through the entire object with negligible interaction with matter (while in the gravitational perturbations case such interaction is self-contained in the equations of motion). Of course, this assumption may at this point seem unjustified, it is nonetheless the only one that we can do without introducing a specific, and at this stage arbitrary, coupling between the object's matter and our test field (see however our comment about absorption below). 
The SV metric with $\ell > 2M$ represents instead a traversable wormhole.
Its throat, differently from a horizon, is traversable in both directions. Since the geometry on the two sides of the wormhole throat is symmetric, we assume that the perturbation will inherit the symmetry of the background. 
This assumption translates into perfect reflection at the throat, which we implement by demanding the perturbation to vanish there, \ie\ $\psi(\ell)= 0$. 

Both the above assumptions can in principle be modified, \eg\ for the limit ultra compact object of Bardeen-like geometries we could introduce an absorption coefficient associated to  the star matter or in the wormhole case we could assume asymmetric stimulation of the wormhole mouth. We leave these extensions of the present study for future investigations.

The direct integration method we used requires an initial guess for the value of the QNM frequency. While in the RBH case we track the mode continuously starting from its ``quasi-Schwarzschild" value obtained for small values of $\ell$, in the ultracompact case because of the discontinuity in the boundary conditions (there is no horizon) and the large values of $\ell$, we do not have any value as a reference to start from. Thus we explored carefully the $(\omega_I,\omega_R)$ plane in order to find the mode with smaller imaginary part, that is the fundamental one.

\section{Results\label{results}}

In what follows we report the QNM spectra for the considered two families of spherically symmetric regular spacetimes.
We focus on the quadrupolar $l=2$ fundamental mode, which is the dominant in the gravitational-wave ringdown signal.
 Note however, that in the ultracompact horizonless cases, these QNMs could become dominant only in the late-time ringdown signal being preceded by a first part of the signal that is very similar to the Schwarzschild one~\cite{Cardoso:2016rao}.

\begin{figure}[ht!]
\centering
\includegraphics[scale=0.35]{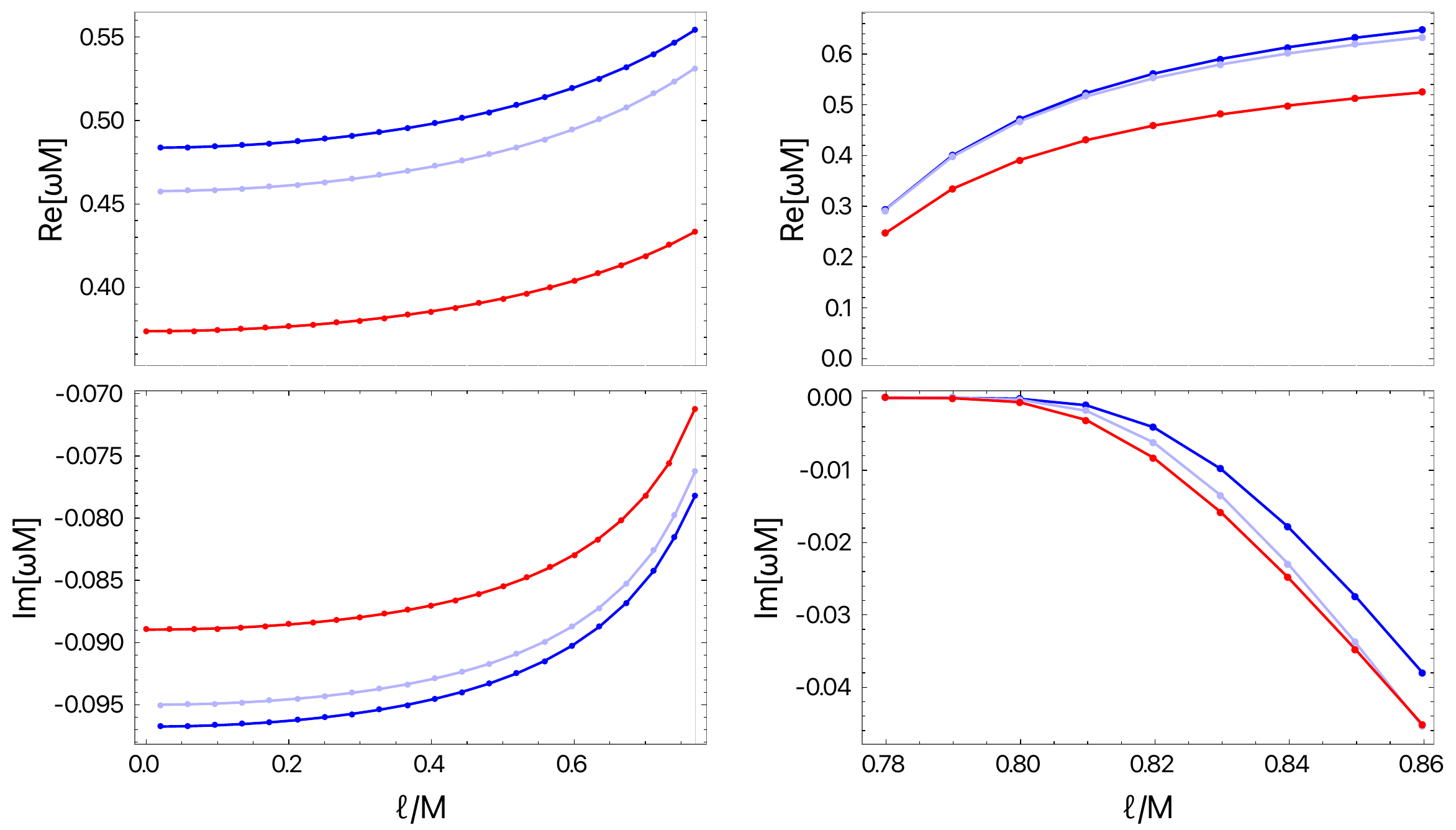}
\caption{Quadrupolar $l=2$ fundamental QNMs of the Bardeen metric for test-field perturbations, $s=0$ (blue), $s=1$ (light purple) and $s=2$ (red). On the left results for values of $\ell$ in the RBH branch that is from $\ell=0$ (Schwarzschild) to $\ell=\ell_\text{ext}=\frac{4}{3\sqrt{3}}M$ (extremal RBH). On the right results for values of $\ell$ in the horizonless branch ($\ell>\ell_\text{ext}$). Note that, in this branch, for values of the regularization parameter near (but not equal to) the extremal one, the imaginary part is extremely small and thus we have very long living modes this is not true for the extremal RBH case, indicated by the vertical line in the left panel.}
\label{Test-Bard}
\end{figure}

\begin{figure}[ht!]
\centering   
\includegraphics[scale=0.35]{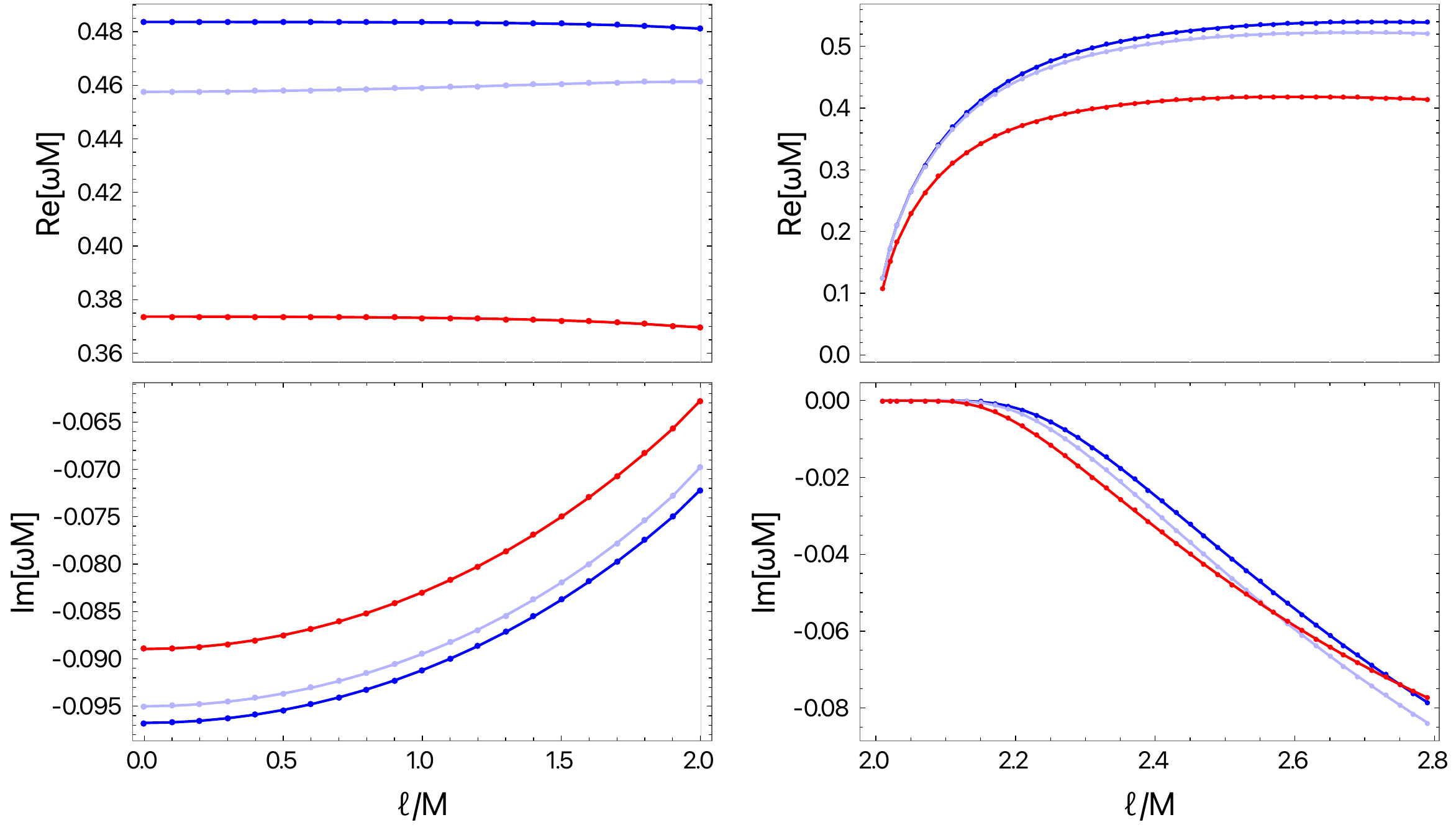}
\caption{Quadrupolar $l=2$ fundamental QNMs of the SV metric for test-field perturbations, $s=0$ (blue), $s=1$ (light purple) and $s=2$ (red). On the left results for values of $\ell$ in the RBH branch, that is from $\ell=0$ (Schwarzschild) to $\ell=2M$ (one-way wormhole with an extremal null throat). On the right results for values of $\ell$ in the horizonless branch ($\ell>2M$). It is worth noticing the relative flatness of the real part curves which highlights weak deviations from the singular GR solution behaviour recovered for $\ell=0$. Note that, in the horizonless branch (right panel), for values of the regularization parameter near (but not equal to) the extremal one, the imaginary part is extremely small and thus we have very long living modes, this is not true for the extremal RBH case, indicated by the vertical line in the left panel.}
\label{Test-SV}
\end{figure}

For test-field perturbations we explore both the RBH and horizonless branches.
For the Bardeen metric we vary the regularization parameter from $\ell=0$, that is Schwarzschild, to roughly the maximum value for which the object still possesses a photon sphere.
In the SV spacetime a photon sphere is always present at the throat and thus there is no upper bound on the value of the regularization parameter, so we let it span in $[0,3.5 M]$.
We show our results in \cref{Test-Bard,Test-SV}. 

Let us note that some results in the test-field approximation were already present in literature, in a specific branch and for specific values of $s$.
Our results are in agreement with those presented \eg\ in Refs.~\cite{Toshmatov:2015wga,Churilova:2019cyt,Franzin:2022iai,Konoplya:2023ahd,Balart:2023odm}.

\begin{figure}[h!]
\centering
\includegraphics[scale=0.43]{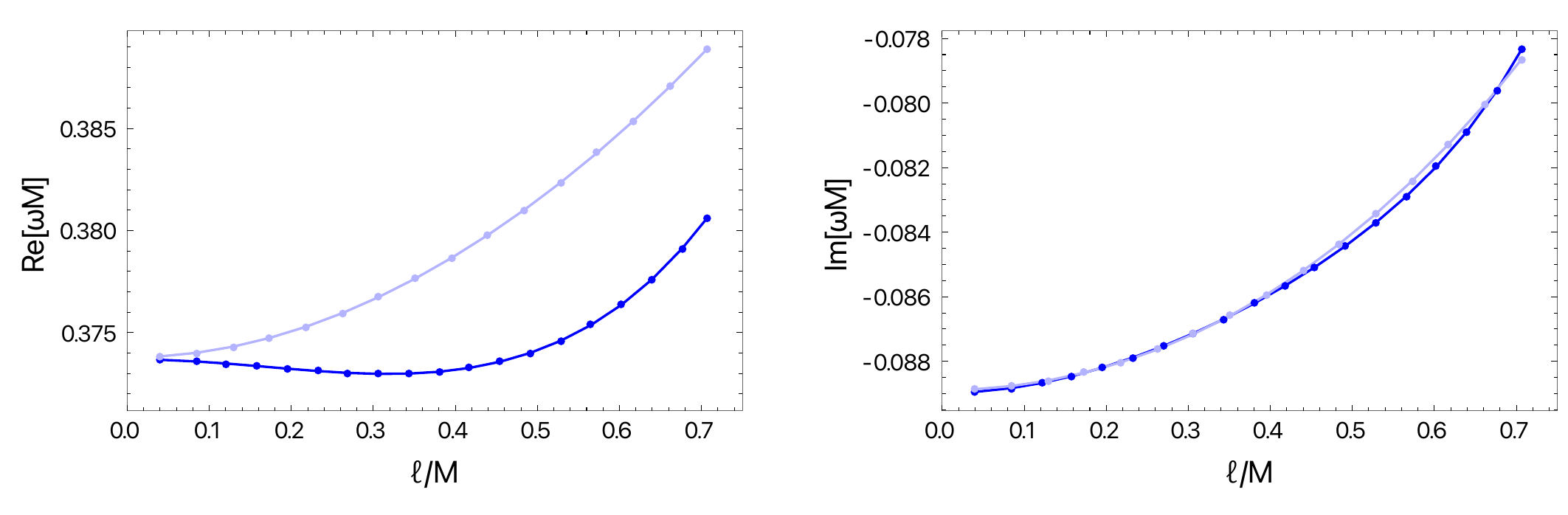}
\caption{Axial (blue) and polar (light purple) $l=2$ gravitational QNMs for the Bardeen metric for values of $\ell$ in the RBH branch that is from $\ell=0$ (Schwarzschild) to $\ell=\ell_\text{ext}=\frac{4}{3\sqrt{3}}M$ (extremal RBH).}
\label{Bard-Grav}
\end{figure}

\begin{figure}[h!]
\centering
\includegraphics[scale=0.43]{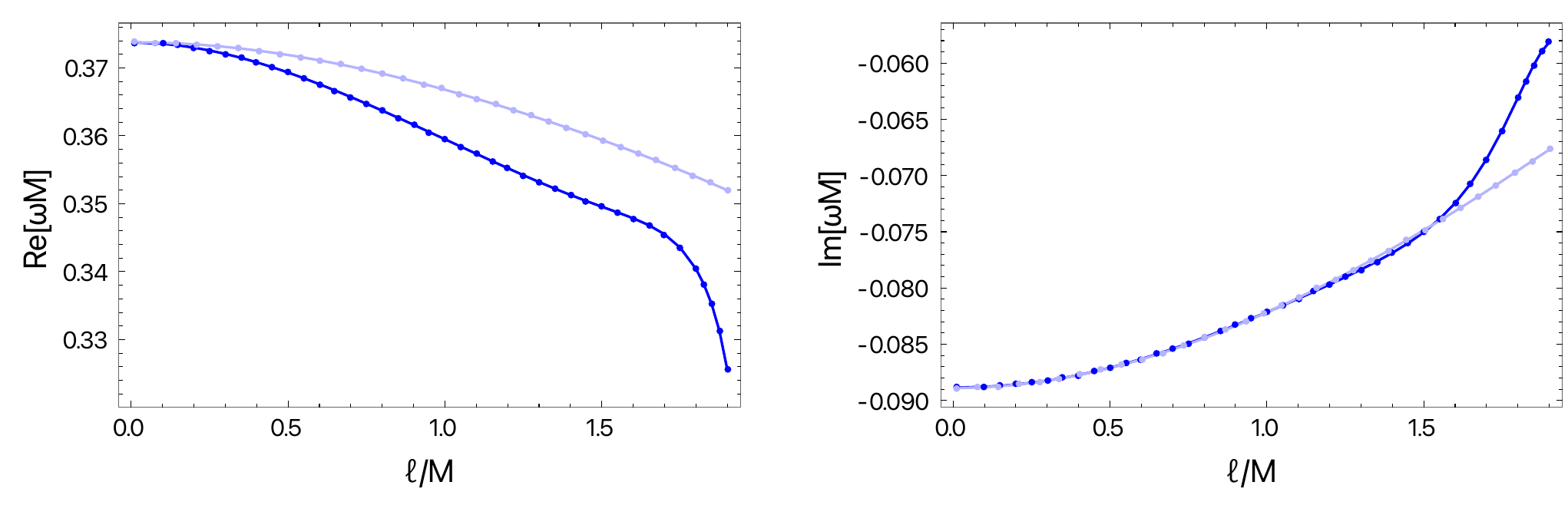}
\caption{Axial (blue) and polar (light purple) $l=2$ gravitational QNMs for the SV metric for values of $\ell$ in the RBH branch that is from $\ell=0$ (Schwarzschild) to $\ell=2M$ (one-way wormhole with an extremal null throat).}
\label{SV-Grav}
\end{figure}

For the full perturbative analysis the computation in the horizonless branch presents some technical difficulties and numerical instabilities, therefore we only report the more solid results for the RBH branch, shown in \cref{Bard-Grav,SV-Grav}.
However, in advance with the discussion in \cref{detectability}, we only need the numerical values of gravitational QNMs in the RBH branch to assess the possible detectability of these deviations with the next generation of gravitational-wave detectors.

\subsection{Summary}

The results for the two families of regular models presents some differences.
For what regards test-field perturbations, the SV spacetime seems to be a better mimicker since, given a certain value of the regularizing parameter $\ell$, its spectrum is more similar to the Schwarzschild one 
(i.e~$|{(\omega_{SV}-\omega_S})/{\omega_S}|< |({\omega_{Bard}-\omega_S})/{\omega_S}|$).
We must say however that, for SV, $\ell$ can span a bigger intervals of values and thus the spectrum can reach higher deviations from Schwarzschild in the imaginary part (some numerical examples are reported in \cref{TabCorr}).
Furthermore the corrections to the real part of the frequency in the RBHs branch are negative (except for $s=1$) while for the Bardeen spacetime are always positive. The reason for this is clear in Schutz-Will WKB approximation \cite{Schutz:1985km} in which $\omega_R \sim V(r_{peak})^{1/2}$ (where $r_{peak}$ is the location of the maximum of the potential). Indeed, compared to the Schwarzschild spacetime, the peak of the potential in the SV spacetime is smaller, whereas in the Bardeen spacetime it is higher. This holds for any spin s of the perturbation except for $s=1$. Indeed, for this value of the spin, in the test-field approximation $V^{\textrm{SV}}=V^{\textrm{Schw}}$ and the small positive corrections in the QNMs of the SV spacetime are only due to the different location of the peak in tortoise coordinates.\\
For what regards full gravitational perturbations instead, the real part of the frequency for SV RBHs presents stronger deviations from the Schwarzschild one in the axial sector. 

For both families of regular models, in the ultracompact branch we found long living modes associated to the trapping of perturbations near the stable photon sphere.
The damping time grows exponentially with the harmonic number and it is longer for values of the regularization parameter near the extremal case, that is for more compact configurations.
This is not surprising, a more efficient trapping is expected in these cases since there is more distance between the two photon spheres and a deeper potential well.
The aforementioned conclusions stand robust within our framework; however, it is crucial to note that they may be influenced by potential interactions between the test field and matter (e.g., through an absorption coefficient). This consideration is particularly significant as the stable photon sphere seats comfortably within the region where the matter stress-energy tensor is non-negligible~\cite{Carballo-Rubio:2022nuj}. 

Finally, we also found that the isospectrality between the axial and the polar sector is broken for both families, mainly in the real part of the frequencies, with deviations that, as expected, are greater for greater values of the regularization parameter.
\begin{table}[ht]
\begin{tabular}[t]{clllllll}
 \multicolumn{8}{c}{Regular Black holes } \\
 \toprule
& \multicolumn{3}{c}{Bardeen} &  & \multicolumn{3}{c}{Simpson--Visser}\\
\midrule
& Test $s\!=\!2$ & \multicolumn{1}{c}{Axial} & \multicolumn{1}{c}{Polar} & \hspace{1em} & Test $s\!=\!2$ &\multicolumn{1}{c}{Axial} & \multicolumn{1}{c}{Polar}\\
\midrule
$\ell/M=0.2\quad$\\
$\Delta_R$ & $\phantom{-}0.0075$ & $-0.0012$ & $\phantom{-}0.0037$ & & $-3\cdot 10^{-5}$ & $-0.0002$ & $-0.0005$ \\
$\Delta_I$ & $\phantom{-}0.0045$ & $\phantom{-}0.0090$ & $\phantom{-}0.0090$ & & $\phantom{-}0.0022$ & $\phantom{-}0.0044$ & $\phantom{-}0.0044$ \\
\midrule
$\ell/M=0.6$\\
$\Delta_R$ & $\phantom{-}0.0808$ & $\phantom{-}0.0069$  &$\phantom{-}0.0297$ & &$-0.0003$ & $-0.0163$ & $-0.0067$ \\
$\Delta_I$ & $\phantom{-}0.0674$ & $\phantom{-}0.0776$ & $\phantom{-}0.0810$ & &$\phantom{-}0.0236$ & $\phantom{-}0.0292$ & $\phantom{-}0.0292$ \\
\midrule
$\ell/M=1.6$\\
$\Delta_R$ & & & & & $-0.0053$ & $-0.0690$ & $-0.0428$ \\
$\Delta_I$ & &  & & & $\phantom{-}0.1798$ & $\phantom{-}0.1854$ & $\phantom{-}0.1776$\\
\bottomrule
\end{tabular}
\qquad
\begin{tabular}[t]{ccc}
 \multicolumn{3}{c}{Horizonless compact objects } \\
 \toprule
& \multicolumn{1}{c}{\quad Bardeen\quad~} & \multicolumn{1}{c}{Simpson--Visser}\\
\midrule
& \multicolumn{1}{c}{Test $s\!=\!2$} & \multicolumn{1}{c}{Test $s\!=\!2$}\\
\midrule
$\delta=0.05$\\
$\Delta_R$ & $\phantom{-}0.1380$ & $-0.1801$ \\
$\Delta_I$ & $\phantom{-}0.9712$ & $\phantom{-}0.9970 $\\
\midrule
$\delta=0.10$\\
$\Delta_R$ & $\phantom{-}0.3613 $ & $ -0.0310$ \\
$\Delta_I$ & $\phantom{-}0.6441$ & $\phantom{-}0.9015$ \\
\midrule
$\delta=0.20$\\
$\Delta_R$ &  & $\phantom{-}0.0482$ \\
$\Delta_I$ &  & $\phantom{-}0.5913$ \\
\bottomrule
\end{tabular}
\caption{Relative deviations from the quadrupolar fundamental Schwarzschild frequency $\Delta_{R/I} = \frac{\omega_{R/I} - \omega^\text{S}_{R/I}}{|\omega^\text{S}_{R/I}|}$ with $\omega^\text{S}M=0.37367-0.08896\iu$, for $s=2$ test-field and linear gravitational perturbations, both in the axial and polar sectors, for selected valued of the regularization parameter.
Results are shown for the Bardeen and SV spacetimes, on the left for the RBH branch and on the right for horizonless configurations.
For the Bardeen metric there are no results for $\ell/M=1.6$ and $\delta=0.2$, with $\delta\equiv\ell/\ell_\text{ext}-1$, since for those values of compactness the  spacetime not only lose the presence of the horizon but even of a photon sphere. For both spacetimes results for axial and polar gravitational perturbations are not reported for horizonless configurations because of the numerical issues present in this branch. Looking at the test field case, it is easy to see the large increment of $\Delta_{I}$ passing from the RBH configurations to the horizonless ones for small $\delta$.}
\label{TabCorr}
\end{table}

\subsection{A connection between the photon sphere instability and the Aretakis instability?}
We found that also in our new class of horizonless ultracompact objects there are long living perturbations modes, associated to the presence of a stable photon sphere. This is usually assumed to be the hint of a non-linear instability.\footnote{ Note however that, as already remarked, our results were obtained by neglecting the possible interactions of the perturbation with matter. Indeed, it can be shown that generically the matter content of these spacetimes is not negligible at the location of the stable photon sphere~\cite{Carballo-Rubio:2022nuj}. If this matter absorbs part of the energy carried by the perturbation, the instability would be most probably tamed.} Indeed the presence of these long living modes in the frequency decomposition is associated with a total perturbation in time domain that decay slower than $1/t$ and this leads to the breaking of linear approximation. 

An intuitive way to see it is the following. In perturbation theory each order is the source of the next one in the linearized Einstein field equations, then if $h^{(n)}$ is the perturbation at the $n$-th order, one has
\be
\Box h^{(2)} \propto h^{(1)} \propto \frac{1}{t^q} \quad\to\quad h^{(2)} \propto \frac{1}{t^{q-2}} 
\ee
thus if $q \leq 1$  then $h^{(2)}$ will be increasing with $t$, so eventually breaking the perturbative order-expansion.

Furthermore, a pseudospectrum analysis~\cite{Boyanov:2022ark} showed that these long living modes can be easily perturbed into unstable modes, \ie\ modes with a positive imaginary part of the frequency. This means that small fluctuations in the system may trigger growing modes and thus lead to an instability. 

It should be noted that, as anticipated, the stable photon sphere responsible for the above mentioned instability is already present in the limiting case of extremal RBHs.
Interestingly, this case is conjectured to be affected by another type of instability, the so-called Aretakis instability~\cite{Aretakis:2011ha,Aretakis:2012ei} which appears to be connected to conserved quantities of extremal horizons. 

Presently, and differently from the photon sphere instability, the Aretakis instability lacks of a sound physical interpretation.
In~\cite{Gralla:2019isj,Ravishankar:2021vip} it has been tentatively connected to the presence of null geodesics trapped near the horizon, that is geodesics that orbit arbitrarily many times around the horizon before falling in. If this connection will be confirmed then it will strongly suggest  that the Aretakis instability should be interpreted as a special case of the photon sphere one. 

However, we have here to notice that the former has been proven to hold also for extremal Kerr BHs~\cite{Aretakis:2012ei} albeit for these BHs the photon sphere at the horizon is actually unstable. Of course, also in this case one can observe geodesics that orbit arbitrarily many times around the horizon before falling in, like it happens around any unstable photon sphere, but, usually, this is not associated to any new instability.

\begin{figure}[h!]
\centering
\includegraphics[width=12 cm]{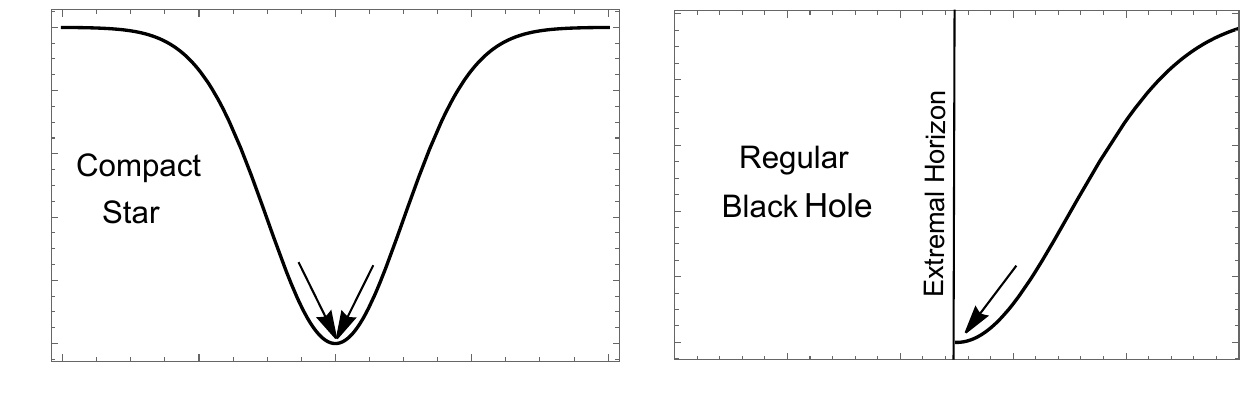}
\caption{Difference between a true stable photon sphere present in the spacetime of compact horizonless objects (left panel) and the ``stable'' photon sphere present at the horizon of extremal BHs and RBHs (right panel). The first one causes a real trapping of modes while in the second case the ``trapped'' modes pass through the horizon in the BH region.}
\label{fig:stable}
\end{figure}

Furthermore, from our previous analysis, it is clear that the damping times for extremal RBHs are of the same order of magnitude of that for sub-extremal ones, while ultracompact objects with stable photon sphere presents very long living modes with damping times several order bigger. This seems to suggest that the photon sphere instability is not triggered or partially suppressed for extremal RBHs. Probably this is due to the fact that an extremal horizon, being an horizon, is not a true stable orbit but can be considered a metastable photon sphere (see \cref{fig:stable}). The presence of an horizon, even if extremal, introduces a source of dissipation: indeed the energy that enters the horizon is completely lost.

In conclusion, there seems to be no ground for a claim that the Aretakis instability is the limit of the instability associated to stable photon spheres for ultra-compact objects when an extremal trapping horizon forms (even when assuming transparent supporting matter for these solutions).

\section{Detectability\label{detectability}}

At this point one may wonder if these QNMs can be distinguished from the QNMs of singular GR BHs in the observed gravitational-wave ringdown signals.
In other words, will we ever be able to prove that the merging objects that produce a given ringdown signal are not singular GR BHs but RBHs? and how many observations we have to combine to do that? 

A general discussion on BH spectroscopy with multiple observations is beyond the scope of this work, a complete review on the topic is~\cite{Berti:2018vdi}, see also~\cite{Yang:2017zxs,Meidam:2014jpa}. However, as a preliminary answer to the above questions, we can here report results obtained within a particular framework for BH spectroscopy, the Parspec framework~\cite{Maselli:2019mjd}. One can take this initial analysis as a proof of principle of detectability of these corrections to BH QNMs. 

\subsection{Parspec framework}

Parspec is an observable-based parametrization of the ringdown signal of rotating BHs beyond GR, it was developed for BH solutions in modified gravity but can be adapted to our phenomenological models of RBHs. We will give here a brief description of this framework.

Let us assume $i = 1, \dots , N$ independent ringdown detections, for which $q$ QNMs are measured.
Each mode of the $i$-th source is parametrized as
\be
\omega_i^{(J)} &:= \Re [\omega_i^{(J)}]=\frac{1}{M_i} \sum_{k=0}^D \chi_i^n \omega_J^{(k)} \left(1+\gamma_i \delta  \omega_J^{(k)}\right),\\
\tau_i^{(J)} &:= \frac{1}{\Im [\omega_i^{(J)}]}=M_i \sum_{k=0}^D \chi_i^n \tau_J^{(k)} \left(1+\gamma_i \delta  \tau_J^{(k)}\right), \label{eq:parspec-expa}
\ee
where $J=1,2,...,q$ labels the mode; $M_i$ and $\chi_i \ll 1$ are the detector-frame mass and spin of the $i$-th source; $D$ is the order of the spin expansion; $\omega_J^{(k)}$ and $t_J^{(k)}$ are the dimensionless coefficients of the spin expansion for a Kerr BH in GR; $\gamma_i$ are dimensionless coupling constants that can depend on the source $i$ but not on the specific QNM $J$, for $\gamma_i \to 0$ the GR BH case is recovered; finally $\delta \omega_J^{(k)}$ and $\delta t_J^{(k)}$ are the ``beyond Kerr'' corrections, in general since all the source dependence is parametrized in $\gamma_i$, these corrections are universal dimensionless numbers.

Since there is no dependence of the corrections on the source, $\gamma_i$ can be set to 1.
We assume perturbative corrections, \ie\ we assume that $\gamma_i \delta \omega^{(k)} \ll 1$ and $\gamma_i \delta t^{(n)} \ll 1$.

It should be noted that  $M_i$ and $\chi_i$ are extracted assuming GR BHs, \ie\ computed from the full inspiral-merger-ringdown waveform within GR. One should extract mass and spin of the BH from the inspiral-merger waveform considering also GR deviations, but this can be very challenging, especially because it requires merger simulations for these RBHs. In this preliminary analysis, we shall assume the shift on the final mass and spin of the source to be negligible. 

To construct the probability distribution of the beyond Kerr parameters we use a Bayesian approach: if we indicate with $ \bar{\theta} $ the parameters (that in our case are $\delta \omega_J^{(k)}$ and $\delta t_J^{(k)}$)  and with $\bar{d}$ a given set of ringdown observations, from the Bayes' theorem we have
\be
P(\bar{\theta}|\bar{d}) \propto \mathcal{L}(\bar{d}| \bar{\theta})P_0(\bar{\theta})
\ee
where $\mathcal{L}(\bar{d}| \bar{\theta})$ is the likelihood function and $P_0(\bar{\theta})$ is the prior on the parameters. Thus from the likelihood we can obtain the full posterior probability distribution $ P( \bar{\theta}|\bar{d}) $  through a Markov chain Monte Carlo (MCMC) method based on the Metropolis-Hastings algorithm.

For each event, the likelihood is chosen to be Gaussian:
\be
\mathcal{L}(\bar{d}| \bar{\theta})= \mathcal{N}(\vec{\mu_i},\Sigma_i)
\ee
where the vector $ \vec{\mu_i}$ is
\be
\vec{\mu_i}=(\vec{\mu_i}^{(1)},...,\vec{\mu_i}^{(q)})^T
\ee
where each $\vec{\mu_i}^{J}$ is a two component vector that depends on the difference between the observed $J=1,...,q$ modes and the parametrized templates in \cref{eq:parspec-expa}:
\be
\vec{\mu}_i^{(J)}= 
\begin{bmatrix}
\omega_i^{(J)}-\omega_{i,\text{obs}}^{(J)}\\
\tau_i^{(J)}-\tau_{i,\text{obs}}^{(J)}\\
\end{bmatrix}
\label{eq:def-mu-parspec} \, ,
\ee
and $\Sigma_i$ is the covariance matrix that includes errors and correlations between the frequencies and damping times of the i-th source.

Since the observed QNMs correspond to different values of $l$ and $m$, \ie\ they are ``quasi-othonormal'', the covariance matrix $\Sigma_i= \diag( \Sigma_i^{(1)},...., \Sigma_i^{(q)}) $ is block-diagonal with each block corresponding to the J-th mode, and thus the likelihood
function can be written as a product of Gaussian distributions:
\be
\mathcal{N}( \vec{\mu_i}, \Sigma_i)= \prod_{J=1}^q \mathcal{N}( \mu_i^{(J)}, \Sigma_i^{(J)})
\ee

Moreover, since we consider N independent detections, the combined likelihood function of the  parameters can be further factorized as:
\be
\mathcal{L}(\vec{d} | \vec{\theta})= \prod_{i=1}^N \mathcal{L}_i(\vec{d} | \vec{\theta})= \prod_{i=1}^N\prod_{J=1}^q \mathcal{N}( \mu_i^{(J)}, \Sigma_i^{(J)})
\ee

\subsection{Results}

We considered only one mode ($l=m=2$) and we stick to 0 order in the spin thus we have:
\be
\omega_i &:= \Re [\omega_i]=\frac{1}{M_i}  \omega^{(0)} \left(1+\delta  \omega^{(0)}\right),\\
\tau_i &:=  \frac{1}{\Im [\omega_i]}=M_i  \tau^{(0)} \left(1+ \delta  \tau^{(0)}\right).
\label{eq:parspec-expa2}
\ee

The analysis can be generalized to higher order in the spin once computed the gravitational QNMs for this rotating RBHs. 

We considered the signal coming from the merger remnant of N binary coalescences as observed by a ground-based 3G detector (ET in the so-called ET-D configuration~\cite{Punturo:2010zz}). The 2N masses of the binary components are drawn from a log-flat distribution between $[5,95]\,M_{\odot}$ and the 2N spins from a uniform distribution between $[-1,1]$. We do not include supermassive BHs in the range of masses since ET will be poorly sensitive to them. We fix the source distance by choosing the signal-to-noise ratio (SNR) of the mode to be $10^2$. We then compute the mass and the spin of the final BH formed after merger using semianalytical relations based on numerical relativity simulations in GR~\cite{Healy:2014yta}. From the final mass of the source we compute the $l=2$ frequency and damping time of a RBHs with that mass, we did the analysis for both Bardeen and SV RBHs. We compute the errors on the modes through a Fisher-matrix approach. 

\begin{figure*}
\centering
\includegraphics[width=0.47\textwidth]{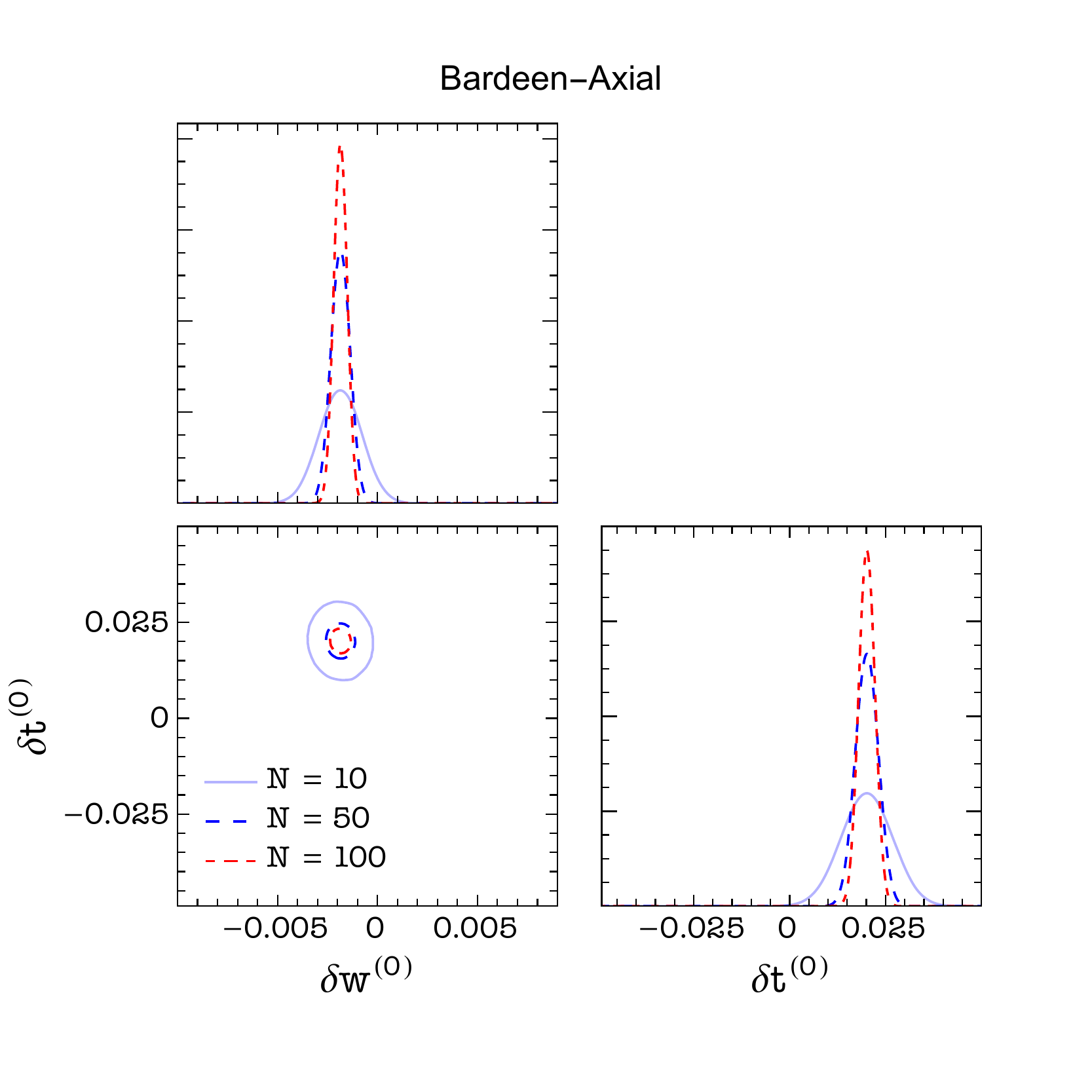}
\qquad
\includegraphics[width=0.47\textwidth]{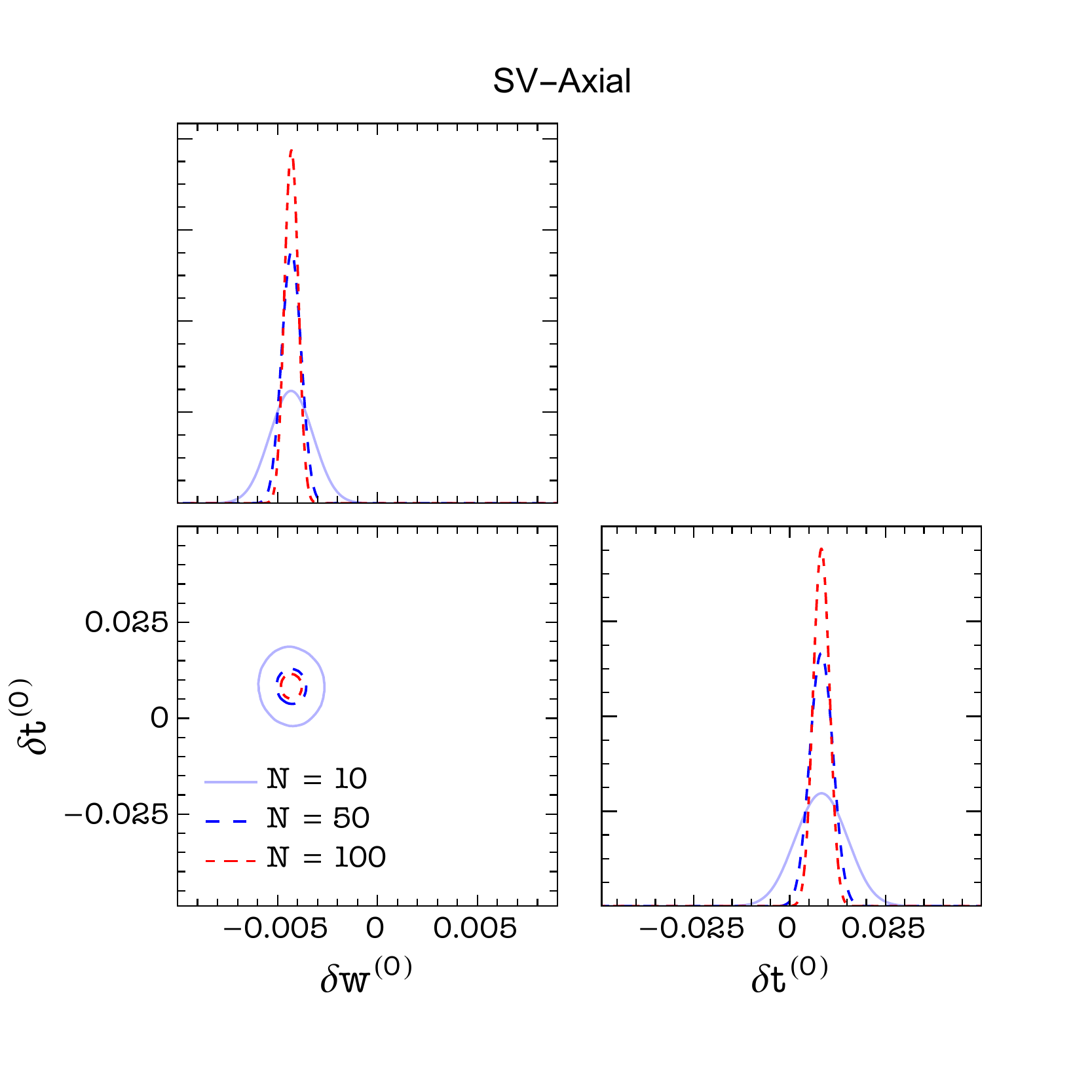}
\qquad
\includegraphics[width=0.47\textwidth]{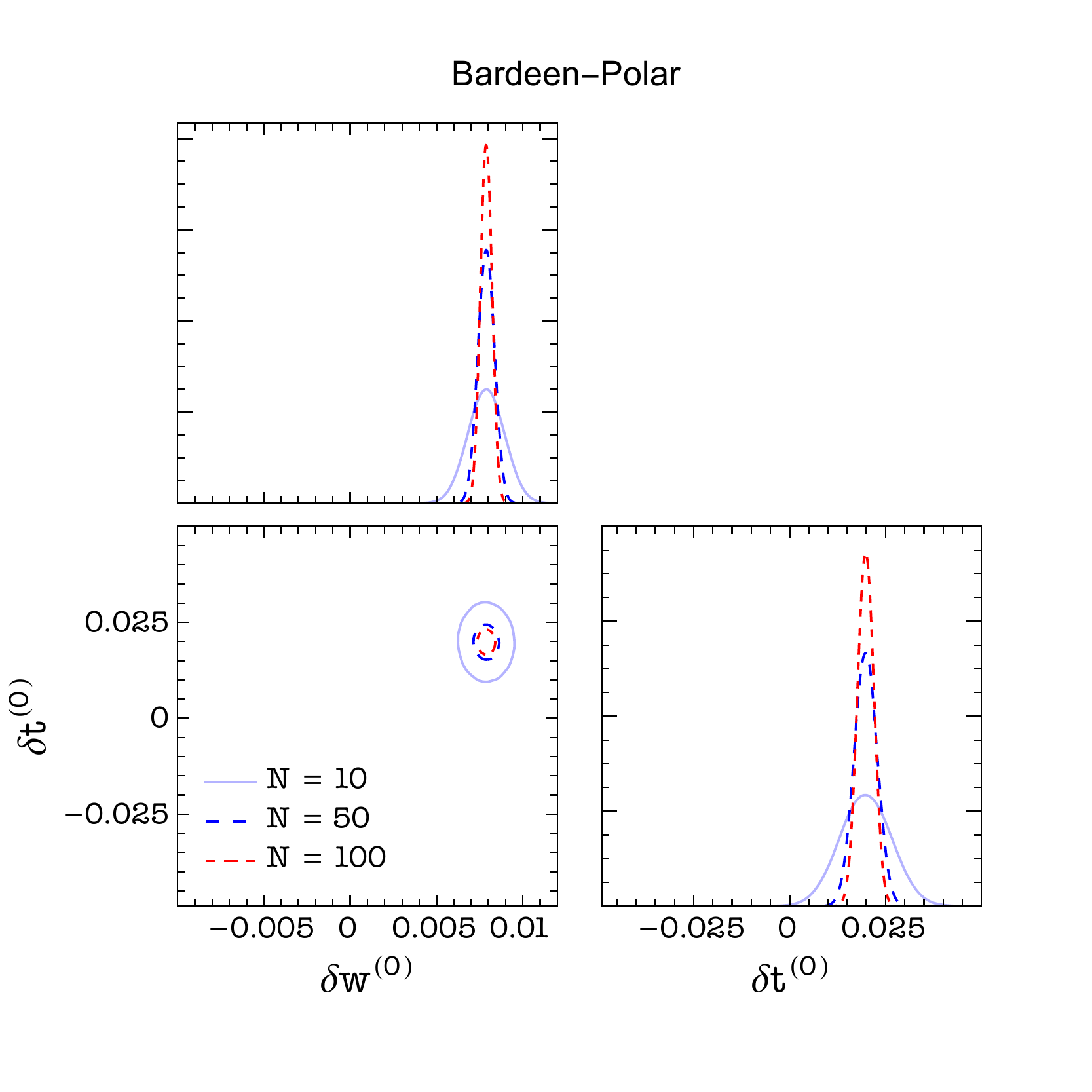}
\qquad
\includegraphics[width=0.47\textwidth]{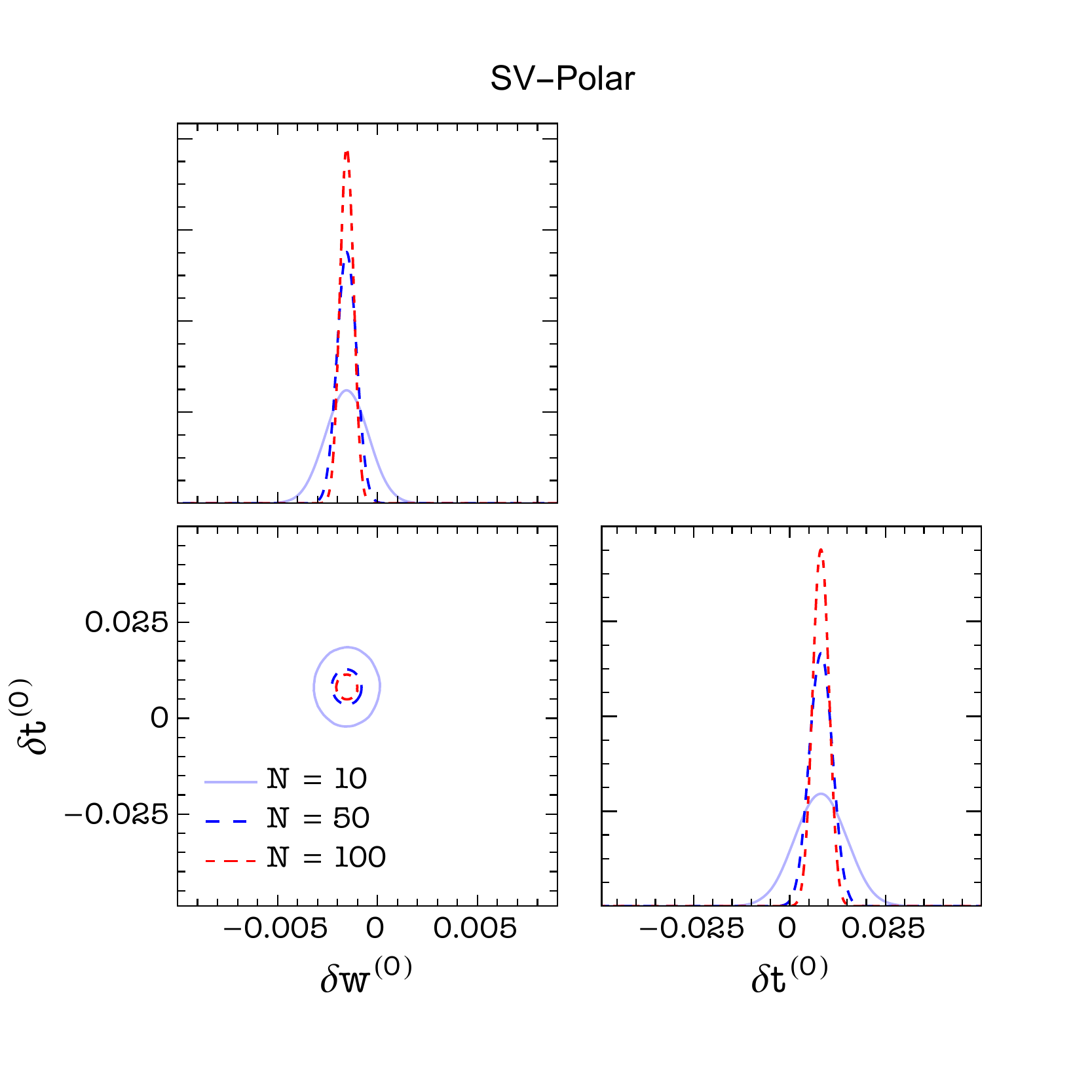}
\caption{Probability distribution functions for corrections to the Schwarzschild $l=m=2$ mode  when we inject as observations axial/polar QNMs of the Bardeen metric with $\ell/M= 0.3 $ (left upper/bottom panel) and axial/polar QNMs of the SV metric with $\ell/M= 0.3 $ (right upper/bottom panel).
Different colors represents results obtained with different numbers $N$ of observed sources.}
\label{fig:detection}
\end{figure*}

Like we can see in \cref{fig:detection}, $\order{100}$ observations with SNR $\approx 10^2$ are enough to exclude with $90 \%$ confidence level the hypotesis of GR singular BHs that is $\delta \omega = \delta \tau = 0$. This even for small, but not Planckian, values of the regularization parameter and for both families of regular models. 

For the Bardeen metric, the strongest constraints come from the real part of the frequency in the polar sector: when $\ell > 0.13 $ its deviation from Schwarzschild real frequency allows to exclude the GR hypothesis at $90\%$ confidence level. For the SV metric the strongest constraints come from the real part of the frequency in the axial sector and to exclude the GR hypothesis we need $\ell \geq 0.19 $. Of course these results depend on the number of sources and their SNR. Here we referred to the case of $\order{100}$ observations with SNR $\sim 10^2$ which should be routinely available with third generation ground-based detectors~\cite{Punturo:2010zz,LIGOScientific:2016wof}.

Note that from the posterior probability distributions is also possible to extract a value for the observed $\delta \omega$ and $\delta \tau$ (with associated errors). This should be the value at which the posterior is peaked. Thus if one knows the dependence of these corrections from the regularization parameter ($\delta \omega(\ell)$ and $\delta \tau (\ell)$) it is also possible to infer the value of $\ell$ from the posterior probability distributions. This dependence could be obtained for example fitting the numerical results for the RBHs QNMs computed in \cref{results}.

\section{Conclusions}

In this paper we have studied test-field and gravitational perturbations on top of the two possible families of spherically symmetric black-hole mimickers, that can be modeled by the Bardeen and SV metric. Both families smoothly interpolate between RBHs and horizonless objects depending on the value of the regularization parameter $\ell$ that enters the metric.

The results for these two families of regular models presents some differences.
For what regards test-field perturbations the SV spacetime seems to be a better mimicker since, given a certain value of the regularizing parameter $\ell$, its spectrum is more similar to the Schwarzschild one (i.e~$|{(\omega_{SV}-\omega_S})/{\omega_S}|< |({\omega_{Bard}-\omega_S})/{\omega_S}|$). We noticed however, that due to the larger span allowed for $\ell$ in the SV case, this can produce higher deviations from Schwarzschild in the imaginary part, as explicitly reported in \cref{TabCorr}.
Furthermore, the corrections to the real part of the frequency in the RBH branch are negative in the SV case while for the Bardeen spacetime are positive.  For what regards full gravitational perturbations instead, the real part of the frequency for SV RBHs presents stronger deviations from the Schwarzschild one in the axial sector.  We also proved that isospectrality between axial and polar QNMs is broken.

For both families of regular models, in the ultracompact branch, we found long living modes whose damping time grows exponentially with the harmonic index $l$ and is longer for values of the regularization parameter near the extremal case, that is for more compact configurations. These modes are associated with the presence of a stable photon sphere in these spacetime and are usually considered a hint for non-linear instability. 

Also the Aretakis instability is expected to affect the extremal RBH case, that is the limiting case between RBHs and horizonless objects. A linear mode analysis is insufficient to confirm it, indeed we find damping times for this case to be of the same order of magnitude of the sub-extremal case. 

In general our analysis demonstrates that there are deviations of the QNM spectrum of these spacetimes from that of a Schwarzschild BH due to the non-zero value of the regularization parameter $\ell$. So, we analysed the possible detectability of these deviations in the observed
gravitational-wave ringdown signals. 
The detectability of such deviations depends on several aspects such as: the number of observations, their SNR and obviously the size of the regularization parameter. Using the Parspec framework for the analysis we showed that these deviations should be detectable in the near future for Bardeen RBHs with $\ell/M > 0.13$ and SV RBHs with $\ell/M> 0.19 $. Indeed with a hundred of observations with SNR $\sim 100$, which should be routinely available with third generation ground-based detectors~\cite{Punturo:2010zz,LIGOScientific:2016wof}, it will be possible to exclude the hypothesis of GR singular BHs with $90\%$ confidence level or to cast constraints on the quantum gravity-induced regularization parameter $\ell$. 
This analysis in only preliminary and we plan to extend it in several ways: using corrections at higher order in the spin, using a more realistic binary population for the sources, and treating also the final mass and spin of the remnant as unknown parameters. 
We also plan to extend this study on gravitational perturbations to rotating BH mimickers and to better investigate the presumed instability of the extremal case. 

In conclusion, in spite of their preliminary nature we do think that the results of the investigations carried on in this work should be taken as a strong encouragement that third generation gravitational-wave experiments have the potential not only to further advance our astrophysical understanding but as well to open a whole new channel into quantum gravity phenomenology.

\begin{acknowledgments}
The authors acknowledge funding from the Italian Ministry of Education and Scientific Research (MIUR) under the grant PRIN MIUR 2017-MB8AEZ\@. This work was supported by the EU Horizon 2020, Research and Innovation Programme under the Marie Sklodowska-Curie Grant Agreement No.\ 101007855.
\end{acknowledgments}

\appendix

\section{Derivation of the perturbative equations\label{app:perturbations}}

To study linear perturbations we start expanding the metric and the matter field around their background values,
\be
g_{\mu \nu} = g^{(0)}_{\mu \nu} + h_{\mu \nu}\,,\quad
A_\mu = A_\mu^{(0)}+\delta A_\mu\,,\quad
\Phi = \Phi^{(0)} + \delta\Phi\,,
\ee
where $g_{\mu\nu}^{(0)}$, $A_\mu^{(0)}$ and $\Phi^{(0)}$ represent the background quantities, while $h_{\mu\nu}$, $\delta A_\mu$ and $\delta\Phi$ are small perturbations.

We further decompose the perturbations in spherical harmonics $Y^{lm}$ and separate them in polar and axial parts according to their parity symmetry, \ie\ $h_{\mu \nu}=$ $h_{\mu \nu}^\text{polar} + h_{\mu \nu}^\text{axial}$, and similarly for the matter field.
In the Regge--Wheeler gauge, $h_{\mu\nu}$ can be written as
\be
h_{\mu \nu}^\text{polar} &=
\sum_{l,m}\begin{pmatrix}
-f(r) H_0^{l m}(t,r) & H_1^{l m}(t,r) & 0 & 0 \\
H_1^{l m}(t,r) & \frac{H_2^{l m}(t,r)}{f(r)}  & 0 & 0 \\
0 & 0 & r^2 K^{l m}(t,r) & 0 \\
0 & 0 & 0 & r^2 \sin^2\theta K^{l m}(t,r)
\end{pmatrix}Y^{l m},
\label{matrixP}\\
h_{\mu \nu}^\text{axial} &=
\sum_{l,m}\begin{pmatrix}
0 & 0 & h_0^{l m}(t,r) S_\theta^{l m} & h_0^{l m}(t,r) S_{\varphi}^{l m} \\
0 & 0 & h_1^{l m}(t,r) S_\theta^{l m} & h_1^{l m}(t,r) S_{\varphi}^{l m} \\
h_0^{l m}(t,r) S_\theta^{l m} & h_1^{l m}(t,r) S_\theta^{l m} & 0 & 0 \\
h_0^{l m}(t,r) S_{\varphi}^{l m} & h_1^{l m}(t,r) S_{\varphi}^{l m} & 0 & 0
\end{pmatrix},
\label{matrixA}
\ee
being $S_b^{l m} \equiv\left(-Y_{,\varphi}^{l m} / \sin\theta, \sin\theta Y_{,\theta}^{l m}\right)$ with $b=\{\theta,\varphi\}$.

Likewise, we expand the electromagnetic potential as
\be
\delta A_\mu^\text{polar} = \sum_{l,m}\left(\frac{u_1^{l m}(t,r)}{r} Y^{l m}, \frac{u_2^{l m}(t,r)}{r f(r)} Y^{l m}, u_3^{l m}(t,r) Y_b^{l m}\right),\quad
\delta A_\mu^\text{axial} = \sum_{l,m}\left(0, 0, u_4^{l m}(t,r) S_b^{l m}\right),
\label{electr}
\ee
being $Y_b^{l m} \equiv\left(Y_{, \theta}^{l m}, Y_{, \varphi}^{l m}\right)$.
Finally, we decompose the scalar perturbation as
\be
\delta\Phi = \sum_{l,m}\frac{\delta\Phi^{l m}}{r}\,Y^{l m}\,.
\ee
In what follows, we drop the symbol $\sum_{l,m}$ and the superscript $lm$ to avoid cluttering the notation. We also assume harmonic time dependence for the perturbation functions, \ie\ for any perturbative quantity $\delta F(t,r)$ we write $\delta F(t,r) = \e^{-\iu\omega t} \delta \tilde{F}(r)$, but we will omit the tilde.

The background metric and scalar field are even under parity transformations, while the background magnetic field is odd.
Hence, to linear order, the axial gravitational perturbations couple to the polar electromagnetic perturbations (Sector I), while the polar gravitational perturbations couple to the axial electromagnetic and the polar scalar perturbations (Sector II).

It is relatively easy to check that the equations derived in the next subsections reproduce well known results in the appropriate limits, \eg\ the Regge--Wheeler--Zerilli gravitational equations for the Schwarzschild spacetime for $\mathcal{L}=0$, $\phi(r)=0$ and $f(r)=1-2M/r$, or those of Ref.~\cite{Nomura:2020tpc} for $\phi(r)=0$.

\subsection{Sector I: axial gravitational--polar electromagnetic}

In this sector the axial gravitational perturbations couple with the polar electromagnetic perturbations.
Let us begin by considering the modified Maxwell equation for a polar electromagnetic perturbation.
At linear order, the field strength squared is unperturbed, $F \approx F^{(0)}$.
It follows that when computing linear perturbations the Lagrangian and its derivatives are unperturbed as well, \eg\ $\mathcal{L}_F \approx \mathcal{L}_{F^{(0)}}$, where $\mathcal{L}_{F^{(0)}} \equiv \p\mathcal{L}/\p F^{(0)}$.
However, to avoid an excessive cluttering of the equations, in what follows we drop the ``$(0)$'' superscript.

The $t$, $r$ and $\theta$ components of the modified Maxwell equation read
\begin{subequations}\be
&f u_1'' + f \left(\phi'-\frac{2 \ell^2 \mathcal{L}_{FF}}{r^5 \mathcal{L}_F}\right) u_1' + \left(\frac{2 \ell^2 f \mathcal{L}_{FF}}{r^6 \mathcal{L}_F}-\frac{r f \phi'+l(l+1)}{r^2}\right) u_1 + \iu \omega u_2' + \iu \omega \left(\frac{1}{r}-\frac{f'}{f}-\frac{2 \ell^2 \mathcal{L}_{FF}}{r^5 \mathcal{L}_F}+\phi'\right) u_2 - \frac{l (l+1) \ell}{r^3}\,h_0 =0\,,\label{sec1max1}\\
&\iu \omega  u_1' - \frac{\iu \omega u_1}{r} + \left(\frac{l (l+1) \e^{-2\phi}}{r^2}-\frac{\omega^2}{f}\right) u_2 + \frac{l (l+1) \ell f \e^{-2\phi}}{r^3}\,h_1 = 0\,,\label{sec1max2}\\
&\iu\omega r \e^{2\phi} u_1 + r f u_2' - f \left(\frac{2 \ell^2 \mathcal{L}_{FF}}{r^4 \mathcal{L}_F} + r \phi' + 1\right) u_2%\nonumber\\
+ \iu \omega  \ell \e^{2 \phi} h_0 + \ell  f^2 h_1' + \ell f \left[f' - f \left(\frac{2 \ell^2 \mathcal{L}_{FF}}{r^5 \mathcal{L}_F}+\phi'+\frac{2}{r}\right)\right] h_1 = 0\,.\label{sec1max3}
\ee\end{subequations}
\Cref{sec1max2} can be solved for $u_2$ and substituting in \cref{sec1max1} gives an equation for $u_1$ with non-homogeneous terms proportional to $h_0$ and $h_1$.
\Cref{sec1max3} is a consequence of the first two equations.

The independent components of the perturbed axial gravitational equations are the $t\theta$, $r\theta$ and $\theta\varphi$
\begin{subequations}\be
&f h_0'' + f \phi' h_0' -\frac{2 r^2 f \left(r \phi'+1\right)+(l-1)(l+2) r^2+4 \ell^2 \mathcal{L}_F}{r^4}\,h_0 + \iu \omega  f h_1' + \frac{\iu \omega f \left(r \phi'+2\right)}{r}\,h_1 - \frac{4 \ell \mathcal{L}_F}{r^3}\,u_1 = 0\,,\label{AxialEq13}\\
&\iu \omega  h_0' - \frac{2 \iu \omega  h_0}{r} + \left(\frac{f \e^{-2 \phi} \left[(l-1)(l+2) r^2+4 \ell^2 \mathcal{L}_F\right]}{r^4}-\omega^2\right)h_1  + \frac{4 \ell \mathcal{L}_F\e^{-2 \phi}}{r^3}\,u_2 = 0\,,\label{AxialEq23}\\
&f \e^{-2 \phi} h_1' + \e^{-2 \phi} \left(f'-f \phi'\right)h_1 +\frac{\iu \omega  h_0}{f} = 0\,.\label{AxialEq34}
\ee\end{subequations}
Solving for $h_0$ in \cref{AxialEq34} and substituting in \cref{AxialEq23} we obtain a dynamical equation for $h_1$, while \cref{AxialEq13} is automatically satisfied as a consequence of the previous equations and the modified Maxwell equations.%
\footnote{We stress again that if we had coupled \eg\ axial gravitational and electromagnetic perturbations as in Refs.~\cite{Ulhoa:2013fca,Dey:2018cws,Toshmatov:2018tyo,Toshmatov:2018ell,Toshmatov:2019gxg,Zhao:2023jiz,Wu:2018xza} we would have found an inconsistency.
Consider for definiteness the gravitational equations that would have been obtained in this case (for brevity we shall omit them here, but they can be easily found in the previously cited literature). First, one would get a $tt$ component implying the decoupling of gravitational and electromagnetic perturbations.
Second, the other independent gravitational equations would provide a system of three coupled equations for $h_0$ and $h_1$.
However, contrary to what happens for \cref{AxialEq34,AxialEq13,AxialEq23}, one of the three equations could be no more deduced from the others, and the only acceptable solution would end up to be the trivial one, \ie\ $h_0=h_1=0$.
A similar argument applies for the modified Maxwell equations.\label{footnote:wrong}}

To make the four independent equations more readable, it is helpful to introduce a new variable $u$, which corresponds to the perturbation of the $tr$ component of the Maxwell tensor,
\be
u \equiv \frac{\iu\omega r u_2 - f \left(u_1 - r u_1'\right)}{r^2 f}\,,\label{udefinition}
\ee
instead of the perturbations of the potential $u_1$ and $u_2$.

Taking the first and second derivative of \cref{udefinition}, solving for $u_1'$ and $u_1''$, and substituting into \cref{sec1max1,sec1max2} we get
\begin{subequations}\be
\mathcal{L}_F u'
+\frac{r^4 \mathcal{L}_F \left(r \phi'+2\right) - 2\ell^2 \mathcal{L}_{FF}}{r^5}\,u
-\frac{l (l+1) \mathcal{L}_F}{r^3 f}\,u_1
-\frac{l (l+1) \ell \mathcal{L}_F}{r^4 f}\,h_0
&= 0\,,\label{newMax1}\\
\iu \omega \mathcal{L}_F u
+\frac{l (l+1) \mathcal{L}_F \e^{-2 \phi}}{r^3}\,u_2
+\frac{l (l+1) \ell f \mathcal{L}_F \e^{-2 \phi}}{r^4}\,h_1 
&= 0\,.\label{newMax2}
\ee\end{subequations}
Solving \cref{newMax1,newMax2,AxialEq34} for $h_0$, $u_1$ and $u_2$ and substituing in \cref{AxialEq23}, we obtain the gravitational dynamical equation for $h_1$
\be
h_1''+
\left(\frac{3 f'}{f}-3 \phi'-\frac{2}{r}\right)h_1'
+\left(\frac{f'^2+\omega^2 \e^{2 \phi}}{f^2}-\frac{l(l+1)-r^2 f''+2 r f' \left(2 r \phi'+1\right)-2}{r^2 f}-\phi''+2 \phi'^2+\frac{2 \phi'}{r}\right)h_1
+\frac{4 \iu \omega \ell \mathcal{L}_F \e^{2 \phi}}{l(l+1) f^2}\,u = 0\,,\label{newEqGravSector1}
\ee
while the electromagnetic dynamical equation for $u$ is obtained from \cref{udefinition} with the substitutions above and using \cref{newEqGravSector1}
\be
u'' + \left(\frac{f'}{f}+\phi'+\frac{4}{r}-\frac{2 \ell^2 \mathcal{L}_{FF}}{r^5 \mathcal{L}_F}\right)u'
+\frac{1}{f^2}\left[\omega^2 \e^{2 \phi} - f \left(\frac{l (l+1) - r f' \left(r \phi'+2\right)}{r^2} + \frac{4 \ell^2 \mathcal{L}_F}{r^4} + \frac{2 \ell^2 \mathcal{L}_{FF} f'}{r^5 \mathcal{L}_F}\right)\right.\nonumber\\
\left.+f^2 \left(\phi'' + \frac{2 r \phi'+2}{r^2} + \frac{6 \ell^2 \mathcal{L}_{FF}}{r^6 \mathcal{L}_F} - \frac{4 \ell^4 \mathcal{L}_{FF}^2}{r^{10} \mathcal{L}_F^2} + \frac{4 \ell^4 \mathcal{L}_{FFF}}{r^{10} \mathcal{L}_F}\right)\right]\,u
-\frac{\iu l (l+1) \left(l^2+l-2\right) \ell \e^{-2 \phi}}{ \omega r^6}\,h_1 = 0\,.\label{newEqEMSector1}
\ee

Finally, \cref{newEqGravSector1,newEqEMSector1} can be written as wave equations by performing the substitutions $h_1 = r\e^\phi \mathcal{A}/f$ and $u = \e^{-\phi}\mathcal{E}/r^2\sqrt{\mathcal{L}_F}$, and by introducing a tortoise-like coordinate $\dd r_*/\dd r=\e^\phi/f$, to get the coupled system
\be
\frac{\dd^2\mathcal{A}}{\dd r_*^2} + \left(\omega^2- V_\mathcal{A} \right)\mathcal{A}
+ \frac{4 \iu \omega  \ell  f \sqrt{\mathcal{L}_F} \e^{-2 \phi}}{l(l+1) r^3}\,\mathcal{E} &= 0\,,\label{GravwaveSec1}\\
\frac{\dd^2\mathcal{E}}{\dd r_*^2} + \left(\omega^2-V_\mathcal{E}\right)\mathcal{E}
-\frac{\iu l (l+1) (l^2+l-2) \ell  f \sqrt{\mathcal{L}_F} \e^{-2 \phi}}{\omega r^3}\,\mathcal{A} &= 0\,,\label{EMwaveSec1}
\ee
where
\be
V_\mathcal{A} &= f \e^{-2 \phi} \left(\frac{f \phi'-f'}{r}+\frac{(l-1)(l+2)+2f}{r^2}\right),\\
V_\mathcal{E} &= f \e^{-2 \phi} \left[\frac{l (l+1)}{r^2}+\frac{4 \ell^2 \mathcal{L}_F}{r^4} + \frac{\ell^2 \mathcal{L}_{FF} f'}{r^5 \mathcal{L}_F}
-\frac{f}{r^6} \left(\frac{\ell^2 \mathcal{L}_{FF} \left(r \phi'+5\right)}{\mathcal{L}_F} - \frac{3 \ell^4 \mathcal{L}_{FF}^2}{r^4 \mathcal{L}_F^2} + \frac{2 \ell^4 \mathcal{L}_{FFF}}{r^4 \mathcal{L}_F}\right)\right].
\ee

\subsection{Sector II: polar gravitational--axial electromagnetic--polar scalar}

In this sector the polar gravitational perturbations couple with the axial electromagnetic and polar scalar perturbations.

Let us begin with the Klein--Gordon equation
\be
\delta\Phi''
-\left(\phi'-\frac{f'}{f}\right)\delta\Phi'
+\left(\frac{\omega^2 \e^{2 \phi}}{f^2}+\frac{\phi'}{r} - \frac{l(l+1) + r f' - r^2 V_{\Phi\Phi}}{r^2 f}\right)\delta\Phi
+\frac{r V_\Phi H_2}{f}
-\left(\frac{\iu\omega r \e^{2 \phi}}{f}\,H_1 - \frac{r(H_0'-H_2'+2K')}{2}\right)\Phi' =0\,.\label{KGsect2_a}
\ee

The field strength squared for an axial perturbation is
\be
F \approx F^{(0)} + \delta F = \frac{\ell^2}{2r^4} - \frac{\ell \e^{-\iu\omega t} \left[\ell K - l(l+1) u_4\right] Y_{lm}}{r^4}\,.
\ee
In this case, when computing linear perturbations to \cref{EqMotMax} we also expand $\mathcal{L}_F$ around $F^{(0)}$, \eg\ $\mathcal{L}_F \approx \mathcal{L}_{F^{(0)}} + \mathcal{L}_{F^{(0)}F^{(0)}}\,\delta F$, and similarly for higher derivatives.

With the further gauge choice $u_3=0$, the $\theta$ component of the modified Maxwell equations is the only non-vanishing, and reads
\be
u_4''
+\left(\frac{f'}{f}-\frac{2 \ell^2 \mathcal{L}_{FF}}{r^5
\mathcal{L}_F}-\phi'\right) u_4'
+\left[\frac{\omega^2 \e^{2 \phi}}{f^2} - \frac{l (l+1)}{r^2 f}\left(1  + \frac{\ell^2 \mathcal{L}_{FF}}{r^4 \mathcal{L}_F}\right)\right]u_4
-\frac{\ell (H_0+H_2)}{2 r^2 f}
+\frac{\ell}{r^2 f}\left(1+\frac{\ell^2 \mathcal{L}_{FF}}{r^4 \mathcal{L}_F}\right)K = 0
\,.\label{Maxwellsector2}
\ee

Lastly, let us consider a polar gravitational perturbation.
The $\theta\varphi$ component of the perturbed gravitational equation requires $H_2=-H_0$.

Using the background equations, the other six independent gravitational equations, namely the $tt$, $tr$, $t\varphi$, $rr$, $r\varphi$ and $\theta\theta$ components, are
\begin{subequations}\be
&f K''
+\left(\frac{f'}{2}+\frac{3 f}{r}\right) K'
-\left(\frac{(l-1)(l+2)}{2 r^2}+\frac{2\ell^2\mathcal{L}_F}{r^4}\right)K
+\frac{f H_0'}{r}
+\frac{2 r f'+2f\left(1-r\phi'\right)+l(l+1)}{2 r^2}\,H_0\nonumber\\
&\qquad+\frac{2 l (l+1)\ell \mathcal{L}_F}{r^4}\,u_4
+\frac{4 f\sqrt{\pi r\phi'}}{r^2}\,\delta\Phi'
+\frac{8\pi r^2 V_\Phi - 4 f\sqrt{\pi r\phi'}}{r^3}\,\delta\Phi = 0\,,\label{PolarGrav11}\\
&K'+
\left(\frac{1}{r}-\frac{f'}{2 f}+\phi'\right)K
+\frac{H_0}{r}
-\frac{\iu l (l+1) H_1}{2 r^2\omega}
+\frac{4 \sqrt{\pi r\phi'}}{r^2}\,\delta\Phi = 0\,,\label{PolarGrav12}\\
&f H_1'
+\left(f'-f\phi'\right)H_1
-\iu\omega H_0
+\iu\omega K
-\frac{4 \iu\omega \ell \mathcal{L}_F}{r^2}\,u_4 = 0\,,\label{PolarGrav14}\\
&\left(\frac{f'}{2}+\frac{f}{r}-f\phi'\right)K'
+\left(\frac{\omega^2 \e^{2\phi}}{f}-\frac{(l-1)(l+2)}{2 r^2}-\frac{2\ell^2\mathcal{L}_F}{r^4}\right)K
+\frac{f H_0'}{r}
-\frac{2 f\left(r\phi'-1\right) - 2 r f' +l(l+1)}{2 r^2}\,H_0
-\frac{2 \iu\omega \e^{2\phi}}{r}\,H_1\nonumber\\
&\qquad+\frac{2 l (l+1)\ell \mathcal{L}_F}{r^4}\,u_4
-\frac{4 f\sqrt{\pi r\phi'}}{r^2}\,\delta\Phi'
+\frac{8\pi r^2 V_\Phi + 4 f\sqrt{\pi r\phi'}}{r^3}\,\delta\Phi = 0\,,\label{PolarGrav22}\\
&H_0'
+\left(\frac{f'}{f}-2\phi'\right)H_0
+K'
-\frac{\iu\omega \e^{2\phi}}{f}\,H_1
-\frac{4\ell \mathcal{L}_F}{r^2}\,u_4'
+\frac{8\sqrt{\pi r\phi'}}{r^2}\,\delta\Phi = 0\,,\label{PolarGrav24}\\
&f H_0''
+\left(2 f'+\frac{2f}{r}-3f\phi'\right)H_0'
+\left(\frac{2\left(r f'-1\right)-2 f\left(r\phi'-1\right)}{r^2}-\frac{\omega^2 \e^{2\phi}}{f}+\frac{4\ell^2\mathcal{L}_F}{r^4}\right)H_0
+f K''
+\left(f'+\frac{2f}{r}-f\phi'\right)K'\nonumber\\
&\qquad+\left(\frac{\omega^2 \e^{2\phi}}{f}+\frac{4\ell^2\mathcal{L}_F}{r^4} + \frac{4\ell^4\mathcal{L}_{FF}}{r^8}\right)K
-2 \iu\omega \e^{2\phi} H_1'
-\frac{\iu\omega \e^{2\phi}\left(r f'+2 f\right)}{r f}\,H_1
-4 l (l+1)\ell\left(\frac{\mathcal{L}_F}{r^4}+\frac{\ell^2\mathcal{L}_{FF}}{r^8}\right)u_4\nonumber\\
&\qquad+\frac{8 f\sqrt{\pi r\phi'}}{r^2}\,\delta\Phi'
+\frac{16\pi r^2 V_\Phi - 8 f\sqrt{\pi r\phi'}}{r^3}\,\delta\Phi
= 0\,.\label{PolarGrav33}
\ee\end{subequations}
The off-diagonal equations are first-order differential equations in the metric perturbations and can be solved for $H_0'$, $H_1'$ and $K'$, hence the $rr$ component \eqref{PolarGrav22} gives an algebraic relation among the metric perturbation functions, which can be used to eliminate $H_0$ from the other equations.

Using these relations as well as the background equations, \cref{PolarGrav11,PolarGrav33} are automatically satisfied.
Let $H_1=\omega R$, then the relevant equations are the $tr$ and $t\varphi$ components, which can be written as a system of two non-homogeneous coupled differential equations
\be
\frac{\dd K}{\dd r} = \alpha_1\,K + \alpha_2\,R + J_1\,,\quad
\frac{\dd R}{\dd r} = \beta_1\,K + \beta_2\,R + J_2\,,\label{KRfirst}
\ee
where
\begin{subequations}\label{alphas}\be
\alpha_1 &= -\frac{2 f r^2 \Big(r\phi'\left[2 f\left(r\phi'+1\right)+\zeta -2 l (l+1)\right]+\zeta -2 l (l+1)+2\Big) - 8 f\ell^2\mathcal{L}_F - l(l+1) r^2[\zeta - l(l+1)]}{2\zeta r^3 f}-\frac{2 r\omega^2 \e^{2\phi}}{\zeta f}\,,\\
\alpha_2 &= \frac{\iu l (l+1)\left[l (l+1)-2 f\left(r\phi'+1\right)\right]}{2\zeta r^2}+\frac{2 \iu\omega^2 \e^{2\phi}}{\zeta}\,,\\
\beta_1 &= -\frac{4 f\left(r^3\phi'\left[f\left(r\phi'+1\right)+\zeta -l (l+1)\right] - [l(l+1)-1] r^2 - 2\ell^2\mathcal{L}_F\right)+r^2 [\zeta -l(l+1)]^2}{2\iu\zeta r^2 f^2}+\frac{2 \iu r^2\omega^2 \e^{2\phi}}{\zeta f^2}\,,\\
\beta_2 &= -\frac{2 r f \phi' [3\zeta +l(l+1)] + 2 f[2\zeta +l(l+1)] + [\zeta -l(l+1)] [2\zeta +l(l+1)]}{2\zeta r f} + \frac{2 r\omega^2 \e^{2\phi}}{\zeta f}\,,\\
J_1 &= -\frac{8\ell f\mathcal{L}_F}{\zeta r^2}\,u_4'
-\frac{4 l (l+1)\ell \mathcal{L}_F}{\zeta r^3}\,u_4
+\frac{8\sqrt{\pi} f\sqrt{r\phi'}}{\zeta r}\,\delta\Phi'
-\frac{16\pi r^2 V_\Phi - 4\sqrt{\pi}\sqrt{r\phi'}\left[2 f\left(r\phi'+2\right)-l (l+1)\right]}{\zeta r^2}\,\delta\Phi\,,\\
J_2 &= \frac{8 \iu\ell \mathcal{L}_F}{\zeta r}\,u_4'
+\frac{4 \iu\ell \left[\zeta +l(l+1)\right]\mathcal{L}_F}{\zeta r^2 f}\,u_4
-\frac{8 \iu\sqrt{\pi}\sqrt{r\phi'}}{\zeta}\,\delta\Phi'
-\frac{16\pi r^2 V_\Phi - 4 \sqrt{\pi} \sqrt{r\phi'}\left[2 f\left(r\phi'+2\right)+\zeta -l(l+1)\right]}{\iu \zeta r f}\,\delta\Phi\,,
\ee\end{subequations}
and $\zeta(r) = r f'-2f (2 r\phi'+1) + l(l+1)$.

Now, the procedure to obtain the equation that governs polar gravitational perturbations follows Zerilli's original derivation.
The task now is to find a new couple of functions $\hat{R}$ and $\hat{H}$ to transform \cref{KRfirst} into
\be
\frac{\dd\hat{K}}{\dd\hat{r}} = \hat{R} + \hat{J}_1\,,\quad
\frac{\dd\hat{R}}{\dd\hat{r}} = -\left(\omega^2-V(r)\right)\hat{K} + \hat{J}_2\,,\label{KRsecond}
\ee
where the new radial variable $\hat{r}$ is given by $\dd\hat{r}/\dd r = 1/n(r)$. To find such transformation we write
\be
K(r) = g_1(r) \hat{K}(\hat{r}) + g_2(r) \hat{R}(\hat{r})\,,\quad
R(r) = k_1(r) \hat{K}(\hat{r}) + k_2(r) \hat{R}(\hat{r})\,,\quad
\frac{\dd\hat{r}}{\dd r} = \frac{1}{n(r)}\,.\label{KRtransf}
\ee
Let us introduce the matricial notation
\be
\bm{\psi} = \begin{pmatrix}K\\R\end{pmatrix}\,,\quad
\bm{A} = \begin{pmatrix}\alpha_1 & \alpha_2\\
\beta_1 & \beta_2\end{pmatrix}\,,\quad
\hat{\bm{\psi}} = \begin{pmatrix}\hat{K}\\\hat{R}\end{pmatrix}\,,\quad
\bm{F} = \begin{pmatrix}g_1 & g_2\\k_1 & k_2\end{pmatrix}\,,\quad
\bm{J} = \begin{pmatrix}J_1\\J_2\end{pmatrix}\,,
\ee
then \cref{KRfirst} can be written as $\dd\bm{\psi}/\dd r = \bm{A}\bm{\psi}+\bm{J}$, \cref{KRtransf} as $\bm{\psi}=\bm{F}\hat{\bm{\psi}}$, which combined with our request \cref{KRsecond} give the system
\be
n \bm{F}^{-1}\left(\bm{A}\bm{F}-\frac{\dd\bm{F}}{\dd r}\right) = 
\begin{pmatrix}0 & 1\\
-\omega^2+V(r) & 0\end{pmatrix}\,,\label{systemZ}
\ee
together with the new source terms $\hat{\bm{J}} = n \bm{F}^{-1}\bm{J}$.

\Cref{systemZ} represents four equations that relate $g_1$, $g_2$, $k_1$, $k_2$, $n$ and $V$ in terms of $\alpha_{1,2}$ and $\beta_{1,2}$.
By equating the coefficients of $\omega^0$ and $\omega^2$ we get eight equations, supplemented by the condition $\det\bm{F}\neq0$, for six unknown functions.
Yet, the system is consistent and admits a solution
\begin{subequations}\be
n(r) &= f \e^{-\phi}\,,\label{eq_n}\\
g_1(r) &= \frac{g_2 \e^{-\phi} [r f'+ l(l+1)] - 2 g_2 f \e^{-\phi} \left(2 r \phi'+1\right)+2 \iu f k_1}{2 r}\,,\\
g_2(r) &= \exp \int\dd r\, \frac{\phi' \left[r f' + f + l(l+1)\right] + r f \phi'' - 2 r f \phi'^2}{\zeta}\,,\label{eq_g2}\\
k_1(r) &= \iu \e^{-\phi}\,\frac{g_2\left[2 f r^2 \left(r\phi'[2 f (r\phi'+1) + \zeta - l(l+1) ] + \zeta - (l-1)(l+2) - 4\ell^2\mathcal{L}_F/r^2\right) + \zeta r^2 [\zeta - l(l+1)]\right] + 2\zeta r^3 f g_2'}{2\zeta r^2 f}\,,\\
k_2(r) &= -\frac{\iu r g_2}{f}\,,\\
V(r) &= \frac{f \e^{-2\phi}}{\zeta^2 r^2}
\Bigg[ 4 f\left(\zeta^2-\zeta (4\lambda +3) + f [2 f+\zeta -4 (\lambda +1)]+2 (\lambda +1)^2+\frac{2\zeta \ell^4\mathcal{L}_{FF}}{r^6}\right) +\zeta \left(\zeta^2-4\zeta (\lambda +1)+8 (\lambda +1)^2\right)\nonumber\\
&\phantom{=} - r^2 f'' \Big(\zeta [\zeta - 4(\lambda+1)]+2 f [4 f+\zeta -4 (\lambda +1)]\Big) +2 r^4 f f''^2 -r\phi' \zeta [\zeta -4 (\lambda +1)] [\zeta -2 (\lambda +1)]\nonumber\\
&\phantom{=} +r\phi' f\left(4 r^2 (\zeta -5 (\lambda +1)) f''+4 f\left(9 r^2 f''-5\zeta +28(\lambda +1)\right)-5\zeta^2+36\zeta (\lambda +1)-72 f^2-40 (\lambda +1)^2\right)\nonumber\\
&\phantom{=} +4 r^2 f\phi'^2\left[f\left(5 r^2 f''+30 f+12\zeta -34 (\lambda +1)\right)+(\zeta -4 (\lambda +1)) (\zeta -3 (\lambda +1))\right] \nonumber\\
&\phantom{=} + 4 r^3 f^2\phi'^3 [44 f+9\zeta -24 (\lambda +1)] + 48 r^4 f^3\phi'^4\nonumber\\
&\phantom{=} + 8\pi r^3\Phi' \left[f\left(13 r^2 f''-26 f-5\zeta +26(\lambda+1)\right)-2\zeta [\zeta -4 (\lambda +1)] + 64 r^2 f^2\phi'^2 + 4 r f\phi' [29 f+5\zeta -16 (\lambda +1)]\right]V_\Phi \nonumber\\
&\phantom{=} +8\pi r^4 \left(42 r f\phi'+ \zeta\right) V_\Phi^2 - 2\zeta r^3 f\phi' V_{\Phi\Phi} \Bigg]\,,
\ee\end{subequations}
where we have introduced $\lambda=(l-1)(l+2)/2$.
\Cref{eq_n} means that the new variable $\hat{r}$ is nothing but the tortoise-like coordinate $r_*$.
The new source terms read
\be
\hat{J}_1 &= \frac{8 f}{\zeta r^2 g_2}\left(\ell \mathcal{L}_F u_4 - \sqrt{\pi r^3\phi'}\delta\Phi\right),\\
\hat{J}_2 &= -\frac{8 f^2 \e^{-\phi}}{\zeta r^2 g_2}\left[
\ell \mathcal{L}_F u_4'
-\frac{\ell \mathcal{L}_F \left[f\left(r^2\left[\zeta -2\lambda +2 r (\zeta -\lambda -1)\phi'\right]-4\ell^2\mathcal{L}_F\right)+r^3 f^2\left(r\phi''+4 r\phi'^2+5\phi'\right)-2\zeta (\lambda +1) r^2\right]}{\zeta r^3 f}\,u_4 \right.\nonumber\\
&\left.-\sqrt{\pi r^3\phi'}\,\delta\Phi'
+\frac{2\pi \zeta r^4 V_\Phi - f\sqrt{\pi r\phi'}\left[r^2\left(\zeta +2\lambda -r [\zeta -2(\lambda+1)]\phi'\right) + 4\ell^2\mathcal{L}_F\right] + r^3 f^2\sqrt{\pi r\phi'}\left(r\phi''+4 r\phi'^2+5\phi'\right)}{\zeta r^2 f}\,\delta\Phi\right].
\ee

The above system and the source terms simplify when the integral in \cref{eq_g2} is given in a closed form, and this depends strongly on the explicit form of the background metric functions.
Remarkably, for $\phi=\phi_0 + \frac{1}{2}\log\left(1-\ell^2/r^2\right)$, with $\phi_0$ being an arbitrary constant, as for the SV spacetime, and for \emph{any} choice for $f$, we find $g_2 = \e^\phi$.
We assume it in what follows.

Finally, combining \cref{KRsecond} we get a master equation for the polar gravitational perturbations coupled with the axial electromagnetic and polar scalar perturbations
\be
\frac{\dd^2\hat{K}}{\dd r_*^2} + \left(\omega^2 - V(r)\right)\hat{K} - n \hat{J}_1' - \hat{J}_2 = 0\,.\label{eqGravSec2}
\ee

The very last step is to use the solutions for the gravitational equations to rewrite the Klein--Gordon and modified Maxwell equations; they read
\be
u_4'' &+ \left(\frac{\zeta -2 (\lambda +1)}{r f}-\frac{2\ell^2\mathcal{L}_{FF}}{r^5\mathcal{L}_F}+3\phi'+\frac{2}{r}\right)u_4'
+\left[\frac{\omega^2 \e^{2\phi}}{f^2}-\left(1+\frac{\ell^2\mathcal{L}_{FF}}{r^4\mathcal{L}_F}\right)\left(\frac{8\ell^2 \mathcal{L}_F}{\zeta r^4} + \frac{2(\lambda +1)}{f r^2}\right)\right] u_4
+\frac{8\ell \sqrt{\pi r\phi'}}{\zeta r^3}\left(1+\frac{\ell^2\mathcal{L}_{FF}}{r^4\mathcal{L}_F}\right)\delta\Phi\0\\
&+\ell\left(1 + \frac{\ell^2\mathcal{L}_{FF}}{r^4\mathcal{L}_F}\right)
\left(\frac{2\lambda - \zeta - 2 r\phi'\left(r f\phi'+f+\zeta -\lambda -1\right)}{\zeta r^3}+\frac{4\ell^2\mathcal{L}_F}{\zeta r^5}+\frac{\lambda+1}{r^3 f}\right)\hat{K}
+\frac{\ell}{r^2}\left(1+\frac{\ell^2\mathcal{L}_{FF}}{r^4\mathcal{L}_F}\right)\hat{K}' = 0\,,\label{eqEMSec2}\\
\delta\Phi'' &+ \left(\frac{2 f+\zeta -2 (\lambda +1)}{r f}+3\phi'\right)\delta\Phi'
+ \left(\frac{\omega^2 \e^{2\phi}}{f^2} + \frac{V_{\Phi\Phi}}{f} - \frac{\zeta}{f r^2} +\frac{r\phi'-2}{r^2}\right)\delta\Phi
-\frac{2\ell \sqrt{r\phi'} \mathcal{L}_F}{\sqrt{\pi} r^2}\,u_4' = 0\label{eqScalSec2}\,.
\ee

\Cref{eqGravSec2,eqEMSec2,eqScalSec2} can be written as wave equations by introducing new variables
\be
\hat{K} = \mathcal{P}\,,\quad
u_4 = f_1 \mathcal{B} + f_2 \mathcal{P} + f_3 \mathcal{S}\,,\quad
\delta\Phi = g_1 \mathcal{S} + g_2 \mathcal{P} + g_3 \mathcal{B}\,,
\ee
where
\be
f_1 &= \frac{c_1}{\sqrt{\mathcal{L}_F}}\,,\quad
f_2 = \frac{\ell }{2 r}\,,\quad
f_3 = \frac{c_3}{\sqrt{\mathcal{L}_F}}\,,\\
g_1 &= \frac{c_3 \ell}{\sqrt{\pi}} \int \dd r\, \frac{\sqrt{\mathcal{L}_F \phi'}}{r^{3/2}} + c_4\,,\quad
g_2 = \frac{\ell^2}{2\sqrt{\pi}} \int \dd r\, \frac{\mathcal{L}_F \sqrt{\phi'}}{r^{5/2}} + c_5\,,\quad
g_3 = \frac{c_1 \ell}{\sqrt{\pi}} \int \dd r\, \frac{\sqrt{\mathcal{L}_F \phi'}}{r^{3/2}} + c_6\,,
\ee
with $c_1c_4-c_3c_6\neq0$, so that
\be
\frac{\dd^2\mathcal{I}}{\dd r_*^2} + \left(\omega^2- V_\mathcal{I} \right)\mathcal{I}
+ \sum_{\mathcal{J\neq I}} c_\mathcal{I,J}\,\mathcal{J} &= 0\,,\label{WaveEqsSec2}
\ee
for $\mathcal{I,J=\{P,B,S\}}$. The potentials $V_\mathcal{I}$ and the coefficients $c_\mathcal{I,J}$ can be given in closed form: %but they are cumbersome and we do not report them here.

\begin{eqnarray}
V_{\hat{K}} &=& \frac{\e^{-2\phi}}{\zeta^2 r^7} \bigg(8 f^2 (\sqrt{\pi} r^{7/2} g_2\sqrt{\phi'} (r^2 (r^2 f'' -2\zeta +4\lambda +r (-9\zeta +10\lambda +10)\phi'+2 )+4\ell^2\mathcal{L}_F )-\ell  f_2 (r^4\mathcal{L}_F (r^2 f''-2\zeta +4\lambda + r (-9\zeta \0\\
&+&10\lambda +10)\phi'+2 )+ 4 r^2\ell^2\mathcal{L}_F^2+2\zeta \ell^2\mathcal{L}_{FF} ) )-8\zeta r^4 f (\ell (-\zeta +4\lambda +4) f_2\mathcal{L}_F +2\sqrt{\pi} r (\zeta -2 (\lambda +1)) g_2\sqrt{r\phi'} )\0\\
&-&16 r^4 f^3 (5 r^2\phi'^2+3 r\phi'+1 ) (\sqrt{\pi} r g_2 \sqrt{r\phi'}-\ell  f_2\mathcal{L}_F )+\zeta^2 r^7 V(r) \e^{2\phi} \bigg)
\end{eqnarray}

\begin{eqnarray}
 c_{\hat{K},u_4} &=& \frac{8 f \e^{-2\phi}}{\zeta^2 r^9}\bigg( f_1 (r^6\ell \mathcal{L}_F (f (-r^2f'' +2 f (r\phi' (5 r\phi'+3 )+1 )+2\zeta -4\lambda +r (9\zeta -10 (\lambda +1))\phi'-2 )+\zeta (\zeta -4 (\lambda +1)) )-4 r^4\ell^3 f\mathcal{L}_F^2\0\\
 &-&2\zeta r^2\ell^3 f\mathcal{L}_{FF} )+\sqrt{\pi} r^{11/2} g_3\sqrt{\phi'} (f (r^2 (r^2 f''-2\zeta +4\lambda +r (-9\zeta +10\lambda +10)\phi'+2 )+4\ell^2\mathcal{L}_F )\0\\
 &-&2 r^2 f^2 (5 r^2\phi'^2+3 r\phi'+1 )-2\zeta r^2 (\zeta -2 (\lambda +1)) )\bigg)
\end{eqnarray}

\begin{eqnarray}
 c_{\hat{K},\delta \Phi} &=&  -\frac{8 f \e^{-2\phi}}{\zeta^2 r^7\sqrt{r\phi'}}\bigg(\ell  g_3\sqrt{r\phi'} (r^4\mathcal{L}_F (f (r^2 f''-2 f (r\phi' (5 r\phi'+3  )+1  )-2\zeta +4\lambda +r (10 (\lambda +1)-9\zeta )\phi'+2  ) \0\\
 &+&\zeta (-\zeta +4\lambda +4)  )+4 r^2\ell^2 f\mathcal{L}_F^2+2\zeta \ell^2 f\mathcal{L}_{FF}  )+\sqrt{\pi} r^4 g_1\phi' (-f (r^2 (r^2 f''-2\zeta +4\lambda +r (-9\zeta +10\lambda +10)\phi'+2  )+4\ell^2\mathcal{L}_F  ) \0\\
 &+&2 r^2 f^2 (5 r^2\phi'^2+3 r\phi'+1  )+2\zeta r^2 (\zeta -2 (\lambda +1))  )\bigg)
\end{eqnarray}

\begin{eqnarray}
 c_{u_4,\hat{K}}&=&\frac{\e^{-2\phi} }{2\sqrt{\pi} r^{11}\zeta^2 (  f_1  g_1-  f_3  g_3)\mathcal{L}_F^2} \bigg(-2 \e^{2\phi}\sqrt{\pi}\zeta^2 (  f_2  g_1-  f_3  g_2)\mathcal{L}_F^2 V(r) r^{11}+4\sqrt{\pi} f^3\mathcal{L}_F  (-8\sqrt{\pi}  f_3  g_2^2\mathcal{L}_F\sqrt{\phi'}  (5 r^2\phi'^2+3 r\phi'+1 ) r^{9/2} \0\\
 &-&8\ell   f_2^2  g_1\mathcal{L}_F^2  (r\phi'  (5 r\phi'+3 )+1 ) r^3+8  f_2  g_2\mathcal{L}_F  (\sqrt{\pi} r\sqrt{r\phi'}  g_1+\ell   f_3\mathcal{L}_F )  (r\phi'  (5 r\phi'+3 )+1 ) r^3+\ell \zeta   g_1  (\mathcal{L}_F r^4 \0\\
 &+&\ell^2\mathcal{L}_{FF} )\phi'  (r\phi'+1 ) ) r^5+ \zeta f\mathcal{L}_F  (2  f_3\mathcal{L}_F  (-16\pi r (\zeta -2 (\lambda +1))\sqrt{r\phi'}  g_2^2+\sqrt{\pi}  (r  (V_{\Phi\Phi} r^2-\zeta )\zeta +8\ell (\zeta -4 (\lambda +1))  f_2\mathcal{L}_F )  g_2 \0\\
 &+&\ell \zeta (\zeta -2 (\lambda +1))  f_2\mathcal{L}_F\sqrt{r\phi'} ) r^4+\sqrt{\pi}  g_1  (-16\ell (\zeta -4 (\lambda +1))  f_2^2\mathcal{L}_F^2 r^4+\mathcal{L}_F  (\ell \zeta (\zeta -4 (\lambda +1))+4 r  f_2  (\lambda \zeta +\zeta \0\\
 &+&8\sqrt{\pi} (\zeta -2 (\lambda +1))  g_2\sqrt{r\phi'} ) ) r^4+\ell^2\zeta (\zeta -4 (\lambda +1)) (\ell -2 r  f_2)\mathcal{L}_{FF} ) ) r^4-f^2  (\sqrt{\pi}  g_1  (-64\ell^3  f_2^2\mathcal{L}_F^4 r^6+8\ell \mathcal{L}_F^3  (r\zeta \ell^2 \0\\
 &+& 2  f_2  (-  f_2  (f'' r^2+(-9\zeta +10\lambda +10)\phi' r-2\zeta +4\lambda +2 ) r^3-\ell \zeta r^2+4\sqrt{\pi}\ell   g_2\sqrt{r^5\phi'} ) ) r^5+\ell^2\zeta \mathcal{L}_F\mathcal{L}_{FF}  (2 r\zeta   f_2  (3 r\phi'+2 )\0\\
 &+&\ell   (-2\zeta +4\lambda +r (-7\zeta +4\lambda +4)\phi'+16\sqrt{\pi}  g_2\sqrt{r\phi'} ) ) r^4+\mathcal{L}_F^2  (  (r\phi'  (\ell \zeta (-7\zeta +4\lambda +4)-16\sqrt{\pi} r (9\zeta -10 (\lambda +1))  f_2  g_2\sqrt{r\phi'} ) \0\\
 &+&2  (8\sqrt{\pi}  g_2  (\ell \sqrt{r\phi'}\zeta +  f_2  (\sqrt{r^7\phi'} f''-2 r (\zeta -2\lambda -1)\sqrt{r\phi'} ) )+\zeta   (\zeta   f_2'' r^3+2\ell (\lambda -\zeta ) ) ) ) r^6+8\ell^3\zeta   (\ell^2-2 r  f_2 (\ell +2 r  f_2) )\mathcal{L}_{FF} ) r^2\0\\
 &+&2\ell^4\zeta^2 (\ell -2 r  f_2)\mathcal{L}_{FF}^2 )-2 r^{7/2}  f_3\mathcal{L}_F^2  (8\pi   g_2^2  (4\mathcal{L}_F\sqrt{\frac{\phi'}{r}}\ell^2+r  ((10 (\lambda +1)-9\zeta )  (r\phi' )^{3/2}-2 (\zeta -2\lambda -1)\sqrt{r\phi'}+\sqrt{r^5\phi'} f'' ) ) r^{9/2}\0\\
 &+&\sqrt{\pi}  g_2  (r^5\zeta^2  (r\phi'-2 )-8  f_2  (\ell \mathcal{L}_F  (f'' r^2+(-9\zeta +10\lambda +10)\phi' r-2\zeta +4\lambda +2 ) r^4+4\ell^3\mathcal{L}_F^2 r^2+2\ell^3\zeta \mathcal{L}_{FF} ) )\sqrt{r}+\zeta^2  (\sqrt{\pi}  g_2'' r^{15/2}\0\\
 &+&\ell \sqrt{\phi'}  (\mathcal{L}_F  (\ell +r  f_2  (3 r\phi'+2 ) ) r^4+\ell^2 (\ell -2 r  f_2)\mathcal{L}_{FF} ) ) ) ) \bigg)
\end{eqnarray}

\begin{eqnarray}
    V_{u_4}&=&-\frac{\e^{-2\phi} f  }{\sqrt{\pi} r^{10}\zeta^2 (  f_1  g_1-  f_3  g_3)\mathcal{L}_F^2}\bigg(  (-16\pi \zeta (\zeta -2 (\lambda +1))  f_2  g_1  g_3\mathcal{L}_F^2\sqrt{r\phi'} r^5+\sqrt{\pi}  f_3\mathcal{L}_F^2  (-\zeta^2 f  g_3'' r^3 \0\\
    &-&  g_3  (-16\sqrt{\pi} r  g_2\sqrt{r\phi'}  (r\phi'  (5 r\phi'+3  )+1  ) f^2+  (\phi'  (\zeta^2+8\sqrt{\pi} (-9\zeta +10\lambda +10)  g_2\sqrt{r\phi'}  ) r^2-2\zeta^2 r+8\sqrt{\pi}  g_2  (4\mathcal{L}_F\sqrt{\frac{\phi'}{r}}\ell^2 \0\\
    &+&\sqrt{r^7\phi'} f''-2 r (\zeta -2\lambda -1)\sqrt{r\phi'}  )  ) f+r\zeta   (  (V_{\Phi\Phi} r^2-\zeta  )\zeta -16\sqrt{\pi} (\zeta -2\lambda -2)  g_2\sqrt{r\phi'}  )  )  ) r^4-16\pi f^2  f_2  g_1  g_3\mathcal{L}_F^2  (\phi'^{3/2}  (5 r\phi'\0\\
    &+&3  ) r^{13/2} +\sqrt{r^{11}\phi'}  )+f  g_1  (\sqrt{\pi}\zeta^2\mathcal{L}_F^2  f_1'' r^7+8\pi \ell \zeta   g_3\mathcal{L}_F  (\mathcal{L}_{FF}\sqrt{r\phi'}\ell^2+\mathcal{L}_F\sqrt{r^9\phi'}  )+8\pi   f_2  g_3\mathcal{L}_F^2  ((-9\zeta +10\lambda +10)\phi'^{3/2} r^{13/2}\0\\
    &+&\sqrt{r^{15}\phi'} f''+4\ell^2\mathcal{L}_F\sqrt{r^7\phi'}-2 (\zeta -2\lambda -1)\sqrt{r^{11}\phi'}  )  )  ) r^3+  f_1  (\ell   f_3  (\zeta^2 (-\zeta +2\lambda +2)\mathcal{L}_F\sqrt{r\phi'} r^7+\zeta^2 f\sqrt{\phi'}  (2\ell^2\mathcal{L}_{FF}\0\\
    &-&r^4\mathcal{L}_F  (3 r\phi'+2  )  ) r^{7/2}-8\sqrt{\pi}  g_2  (\mathcal{L}_F  (\zeta (\zeta -4 (\lambda +1))+f  (-f'' r^2+(9\zeta -10 (\lambda +1))\phi' r+2\zeta -4\lambda +2 f  (r\phi'  (5 r\phi'+3  )+1  )-2  )  ) r^4\0\\
    &-&4\ell^2 f\mathcal{L}_F^2 r^2-2\ell^2\zeta f\mathcal{L}_{FF}  ) r^3  )\mathcal{L}_F^2+\sqrt{\pi}  g_1  (8\ell \mathcal{L}_F^3  (2 r  f_2  (r\phi'  (5 r\phi'+3  )+1  ) f^2-  (\ell \zeta +r  f_2  (-2\zeta +4\lambda +r  ((-9\zeta +10\lambda +10)\phi'+r f''  )\0\\
    &+&2  )  ) f+r\zeta (\zeta -4 (\lambda +1))  f_2  ) r^6-32\ell^3 f  f_2\mathcal{L}_F^4 r^5+\ell^2\zeta^2\mathcal{L}_F\mathcal{L}_{FF}  (\zeta -4 (\lambda +1)+f  (3 r\phi'+2  )  ) r^4-2\zeta \mathcal{L}_F^2  (\zeta (\lambda +1) r^6 \0\\
    &+&4\ell^3 f (\ell +2 r  f_2)\mathcal{L}_{FF}  ) r^2-2\ell^4\zeta^2 f\mathcal{L}_{FF}^2  )  ) \bigg)
\end{eqnarray}

\begin{eqnarray}
    c_{u_4, \delta \Phi} &=&\frac{\e^{-2\phi} f }{\sqrt{\pi} r^{10}\zeta^2 (  f_1    g_1-    f_3    g_3)\mathcal{L}_F^2}\bigg(8\pi   f_2    g_1^2\mathcal{L}_F^2 (2 f^2\sqrt{r\phi'} (r\phi' (5 r\phi'+3  )+1  ) r^2+2\zeta (\zeta -2 (\lambda +1))\sqrt{r\phi'} r^2-f (- ((\zeta -2\lambda )\sqrt{\phi'} r^{5/2}  )\0\\
    &+&(-9\zeta +10\lambda +10) (r\phi'  )^{3/2} r^2+\sqrt{r\phi'} ( (f'' r^2-\zeta +2\lambda +2  ) r^2+4\ell^2\mathcal{L}_F  )  )  ) r^6+\ell     f_3^2\mathcal{L}_F^2 (\sqrt{r\phi'} (r^4\mathcal{L}_F (\zeta -2\lambda +f (3 r\phi'+2  )-2  )\0\\
    &-&2\ell^2 f\mathcal{L}_{FF}  )\zeta^2+8\sqrt{\pi}    g_2 (\mathcal{L}_F (\zeta (\zeta -4 (\lambda +1))+f (-f'' r^2+(9\zeta -10 (\lambda +1))\phi' r+2\zeta -4\lambda +2 f (r\phi' (5 r\phi'+3  )+1  )-2  )  ) r^4\0\\
    &-&4\ell^2 f\mathcal{L}_F^2 r^2-2\ell^2\zeta f\mathcal{L}_{FF}  )  ) r^3-\sqrt{\pi}\zeta f    g_1\mathcal{L}_F (\zeta \mathcal{L}_F    f_3'' r^7+8\sqrt{\pi}\ell     g_1 (\mathcal{L}_F r^4+\ell^2\mathcal{L}_{FF}  )\sqrt{r\phi'}  ) r^3+\sqrt{\pi}    f_3 (\zeta^2 f\mathcal{L}_F^2    g_1'' r^{10} \0\\
    &+&   g_1 (8\ell \mathcal{L}_F^3 (-2 r  f_2 (r\phi' (5 r\phi'+3  )+1  ) f^2+ (\ell  (\zeta +4\sqrt{\pi}    g_2\sqrt{r\phi'}  )+r  f_2 (-2\zeta +4\lambda +r ((-9\zeta +10\lambda +10)\phi'+r f''  )+2  )  ) f \0\\
    &+&r\zeta (-\zeta +4\lambda +4)  f_2  ) r^6+32\ell^3 f  f_2\mathcal{L}_F^4 r^5-\ell^2\zeta^2\mathcal{L}_F\mathcal{L}_{FF} (\zeta -4 (\lambda +1)+f (3 r\phi'+2  )  ) r^4+\mathcal{L}_F^2 (\zeta  (\zeta  (V_{\Phi\Phi} r^2-\zeta +2\lambda +2  )\0\\
    &-&16\sqrt{\pi} (\zeta -2 (\lambda +1))    g_2\sqrt{r\phi'}  ) r^6-16\sqrt{\pi} f^2    g_2\sqrt{r\phi'} (r\phi' (5 r\phi'+3  )+1  ) r^6+f ( (-2\zeta^2+r\phi' (\zeta^2+8\sqrt{\pi} (10 (\lambda +1)\0\\
    &-&9\zeta )    g_2\sqrt{r\phi'}  )+8\sqrt{\pi}    g_2 (\sqrt{r^5\phi'} f''-2 (\zeta -2\lambda -1)\sqrt{r\phi'}  )  ) r^6+8\ell^3\zeta (\ell +2 r  f_2)\mathcal{L}_{FF}  )  ) r^2+2\ell^4\zeta^2 f\mathcal{L}_{FF}^2  )  )\bigg)
\end{eqnarray}

\begin{eqnarray}
c_{\delta \Phi, \hat{K}} &=&  \frac{\e^{-2\phi}}{2\sqrt{\pi} r^{11}\zeta^2 ( f_1 g_1- f_3 g_3)\mathcal{L}_F^2}\bigg(-2 \e^{2\phi}\sqrt{\pi}\zeta^2 ( f_1 g_2- f_2 g_3)\mathcal{L}_F^2 V(r) r^{11}-4\sqrt{\pi} f^3\mathcal{L}_F  ( g_3  (8\sqrt{\pi} f_2 g_2\mathcal{L}_F\sqrt{r\phi'}  (r\phi'  (5 r\phi'+3  )+1  ) r^4 \0\\
&-&8\ell  f_2^2\mathcal{L}_F^2  (r\phi'  (5 r\phi'+3  )+1  ) r^3+\ell \zeta   (\mathcal{L}_F r^4+\ell^2\mathcal{L}_{FF}  )\phi'  (r\phi'+1  )  )-8 r^3 f_1 g_2\mathcal{L}_F  (\sqrt{\pi} g_2\sqrt{r^3\phi'}-\ell  f_2\mathcal{L}_F  )  (r\phi'  (5 r\phi'+3  )+1  )  ) r^5\0\\
&+&\zeta f\mathcal{L}_F  (2 r^4 f_1\mathcal{L}_F  (16\pi r (\zeta -2\lambda -2)\sqrt{r\phi'} g_2^2+\sqrt{\pi}  (r\zeta   (\zeta -V_{\Phi\Phi} r^2  )-8\ell (\zeta -4\lambda -4) f_2\mathcal{L}_F  ) g_2+\ell \zeta (-\zeta +2\lambda +2) f_2\mathcal{L}_F\sqrt{r\phi'}  )\0\\
&-&\sqrt{\pi} g_3  (-16\ell (\zeta -4\lambda -4) f_2^2\mathcal{L}_F^2 r^4+\mathcal{L}_F  (\ell \zeta (\zeta -4\lambda -4)+4 r f_2  (\lambda \zeta +\zeta +8\sqrt{\pi} (\zeta -2\lambda -2) g_2\sqrt{r\phi'}  )  ) r^4 \0\\
&+&\ell^2\zeta (\zeta -4\lambda -4) (\ell -2 r f_2)\mathcal{L}_{FF}  )  ) r^4+f^2  (\sqrt{\pi} g_3  (-64\ell^3 f_2^2\mathcal{L}_F^4 r^6+8\ell \mathcal{L}_F^3  (r\zeta \ell^2+2 f_2  (- f_2  (-2\zeta +4\lambda +r  ((-9\zeta +10\lambda +10)\phi' \0\\
&+&r f''  )+2  ) r^3-\ell \zeta r^2+4\sqrt{\pi}\ell  g_2\sqrt{r^5\phi'}  )  ) r^5+\ell^2\zeta \mathcal{L}_F\mathcal{L}_{FF}  (2 r\zeta  f_2  (3 r\phi'+2  )+\ell   (-2\zeta +4\lambda +r (-7\zeta +4\lambda +4)\phi'\0\\
&+&16\sqrt{\pi} g_2\sqrt{r\phi'}  )  ) r^4+\mathcal{L}_F^2  (  (r\phi'  (\ell \zeta (-7\zeta +4\lambda +4)+16\sqrt{\pi} r (-9\zeta +10\lambda +10) f_2 g_2\sqrt{r\phi'}  )+2  (8\sqrt{\pi} g_2  (\ell \sqrt{r\phi'}\zeta \0\\
&+& f_2  (\sqrt{r^7\phi'} f''-2 r (\zeta -2\lambda -1)\sqrt{r\phi'}  )  )+\zeta   (\zeta  f_2'' r^3-2\ell \zeta +2\ell \lambda  )  )  ) r^6+8\ell^3\zeta   (\ell^2-2 r f_2 (\ell +2 r f_2)  )\mathcal{L}_{FF}  ) r^2\0\\
&+&2\ell^4\zeta^2 (\ell -2 r f_2)\mathcal{L}_{FF}^2  )-2 r^{7/2} f_1\mathcal{L}_F^2  (8\pi  g_2^2  (4\mathcal{L}_F\sqrt{\frac{\phi'}{r}}\ell^2+r  ((-9\zeta +10\lambda +10)  (r\phi'  )^{3/2}-2 (\zeta -2\lambda -1)\sqrt{r\phi'}\0\\
&+&\sqrt{r^5\phi'} f''  )  ) r^{9/2}+\sqrt{\pi} g_2  (r^5\zeta^2  (r\phi'-2  )-8\ell  f_2  (\mathcal{L}_F  (-2\zeta +4\lambda +r  ((-9\zeta +10\lambda +10)\phi'+r f''  )+2  ) r^4+4\ell^2\mathcal{L}_F^2 r^2\0\\
&+&2\ell^2\zeta \mathcal{L}_{FF}  )  )\sqrt{r}+\zeta^2  (\sqrt{\pi} g_2'' r^{15/2}+\ell \sqrt{\phi'}  (\mathcal{L}_F  (\ell +r f_2  (3 r\phi'+2  )  ) r^4+\ell^2 (\ell -2 r f_2)\mathcal{L}_{FF}  )  )  )  )\bigg)
\end{eqnarray}
 
\begin{eqnarray}
    c_{\delta \Phi, u_4} &=& -\frac{\e^{-2\phi} f}{\sqrt{\pi} r^{10}\zeta^2 ( f_1 g_1- f_3 g_3)\mathcal{L}_F^2}\bigg(\ell  f_1^2\mathcal{L}_F^2 (\zeta^2\mathcal{L}_F\sqrt{r\phi'} (\zeta -2 (\lambda +1)+f (3 r\phi'+2 ) ) r^4+8\sqrt{\pi} g_2 (\mathcal{L}_F (\zeta (\zeta -4 (\lambda +1))\0\\
    &+&f (-f'' r^2+(9\zeta -10 (\lambda +1))\phi' r+2\zeta -4\lambda +2 f (r\phi' (5 r\phi'+3 )+1 )-2 ) ) r^4-4\ell^2 f\mathcal{L}_F^2 r^2-2\ell^2\zeta f\mathcal{L}_{FF} )\0\\
    &-&2\ell^2\zeta^2 f\mathcal{L}_{FF}\sqrt{r\phi'} ) r^3+\sqrt{\pi} g_3\mathcal{L}_F (-8\sqrt{\pi} f_2 g_3\mathcal{L}_F (-2 (5\phi'^{5/2} r^{15/2}+3\phi'^{3/2} r^{13/2}+\sqrt{r^{11}\phi'} ) f^2+ ((-9\zeta +10\lambda +10)\phi'^{3/2} r^{13/2}\0\\
    &+&\sqrt{r^{15}\phi'} f''+4\ell^2\mathcal{L}_F\sqrt{r^7\phi'}-2\zeta \sqrt{r^{11}\phi'}+4\lambda \sqrt{r^{11}\phi'}+2\sqrt{r^{11}\phi'} ) f-2\zeta (\zeta -2 (\lambda +1))\sqrt{r^{11}\phi'} )-\zeta f (\zeta \mathcal{L}_F f_1'' r^7 \0\\
    &+&8\sqrt{\pi}\ell  g_3 (\mathcal{L}_{FF}\sqrt{r\phi'}\ell^2+\mathcal{L}_F\sqrt{r^9\phi'} ) ) ) r^3+\sqrt{\pi} f_1 (\zeta^2 f\mathcal{L}_F^2 g_3'' r^{10}+ g_3 (-8\ell \mathcal{L}_F^3 (2 r f_2 (r\phi' (5 r\phi'+3 )+1 ) f^2\0\\
    &-& (\ell  (\zeta +4\sqrt{\pi} g_2\sqrt{r\phi'} )+r f_2 (f'' r^2+(-9\zeta +10\lambda +10)\phi' r-2\zeta +4\lambda +2 ) ) f+r\zeta (\zeta -4 (\lambda +1)) f_2 ) r^6+32\ell^3 f f_2\mathcal{L}_F^4 r^5\0\\
    &-&\ell^2\zeta^2\mathcal{L}_F\mathcal{L}_{FF} (\zeta -4 (\lambda +1)+f (3 r\phi'+2 ) ) r^4+\mathcal{L}_F^2 (-16\sqrt{\pi} r g_2 (5\phi'^{5/2} r^{15/2}+3\phi'^{3/2} r^{13/2}+\sqrt{r^{11}\phi'} ) f^2\0\\
    &+& (-2\zeta^2 r^6+\phi' (\zeta^2 r^6+8\sqrt{\pi} g_2 ((7\zeta -6 (\lambda +1))\sqrt{r\phi'} r^5+16 (-\zeta +\lambda +1)\sqrt{r^{11}\phi'} ) r ) r-8\sqrt{\pi} g_2 (2 (\zeta -2\lambda -1)\sqrt{r^{11}\phi'}\0\\
    &-&\sqrt{r^{15}\phi'} f'' ) r+8\ell^3\zeta (\ell +2 r f_2)\mathcal{L}_{FF} ) f+r\zeta  (r^5\zeta  (V_{\Phi\Phi} r^2-\zeta +2\lambda +2 )-16\sqrt{\pi} (\zeta -2 (\lambda +1)) g_2\sqrt{r^{11}\phi'} ) ) r^2\0\\
    &+&2\ell^4\zeta^2 f\mathcal{L}_{FF}^2 ) )\bigg)
\end{eqnarray}

\begin{eqnarray}
    V_{\delta \Phi} &=& \frac{\e^{-2\phi}}{\sqrt{\pi} r^{13}\zeta^2 (f_1g_1-f_3g_3)\mathcal{L}_F^2} \bigg(\sqrt{\pi} rg_3 (f\mathcal{L}_F (-16\sqrt{\pi} f^2f_2g_1\mathcal{L}_F\sqrt{\phi'} (r\phi' (5 r\phi'+3 )+1 ) r^5-16\sqrt{\pi}\zeta (\zeta -2 (\lambda +1))f_2g_1\mathcal{L}_F\sqrt{\phi'} r^5 \0\\ 
    &+&f (\zeta^2\mathcal{L}_Ff_3'' r^{13/2}+8\sqrt{\pi}g_1\sqrt{\phi'} (\mathcal{L}_F (\ell \zeta +rf_2 (-2\zeta +4\lambda +r ((-9\zeta +10\lambda +10)\phi'+r f'' )+2 ) ) r^4+4\ell^2f_2\mathcal{L}_F^2 r^3+\ell^3\zeta \mathcal{L}_{FF} ) ) ) r^{11/2}\0\\
    &+&f_3 (8\ell f\mathcal{L}_F^3 (2 rf_2 (r\phi' (5 r\phi'+3 )+1 ) f^2- (\ell \zeta +rf_2 (-2\zeta +4\lambda +r ((-9\zeta +10\lambda +10)\phi'+r f'' )+2 ) ) f+r\zeta (\zeta -4\lambda -4)f_2 ) r^6 \0\\
    &-&32\ell^3 f^2f_2\mathcal{L}_F^4 r^5+\ell^2\zeta^2 f\mathcal{L}_F\mathcal{L}_{FF} (\zeta -4\lambda +f (3 r\phi'+2 )-4 ) r^4+2\zeta \mathcal{L}_F^2 (-\zeta (\lambda +1) f r^6-4\ell^3 f^2 (\ell +2 rf_2)\mathcal{L}_{FF} ) r^2\0\\
    &-&2\ell^4\zeta^2 f^2\mathcal{L}_{FF}^2 ) r^2 )-f_1\mathcal{L}_F^2 (rg_1 (-16\pi f^3g_2\sqrt{\phi'} (r\phi' (5 r\phi'+3 )+1 ) r^{21/2}+\sqrt{\pi}\zeta f ( (V_{\Phi\Phi} r^2-\zeta )\zeta \0\\
    &-&16\sqrt{\pi} (\zeta -2\lambda -2)g_2\sqrt{r\phi'} ) r^{10}+f^2 (\sqrt{\pi}\zeta^2 (r\phi'-2 ) r^{10}+8\pi g_2\sqrt{\phi'} ( (-2\zeta +4\lambda +r ((-9\zeta +10\lambda +10)\phi'+r f'' )+2 ) r^2 \0\\
    &+&4\ell^2\mathcal{L}_F ) r^{17/2} ) )+f (\sqrt{\pi}\zeta^2 fg_1'' r^{13}+\ell f_3 ( (r^{9/2}\mathcal{L}_F\sqrt{\phi'} (\zeta -2\lambda +f (3 r\phi'+2 )-2 )-2\ell^2 f\mathcal{L}_{FF}\sqrt{r\phi'} )\zeta^2\0\\
    &+&8\sqrt{\pi}g_2 (\mathcal{L}_F (\zeta (\zeta -4 (\lambda +1))+f (-f'' r^2+(9\zeta -10 (\lambda +1))\phi' r+2\zeta -4\lambda +2 f (r\phi' (5 r\phi'+3 )+1 )-2 ) ) r^4\0\\
    &-&4\ell^2 f\mathcal{L}_F^2 r^2-2\ell^2\zeta f\mathcal{L}_{FF} ) ) r^6 ) ) \bigg)
\end{eqnarray}

\subsection{Static perturbations\label{app:static}}

Static perturbations are easier to derive and we report them for completeness.

\medskip
In sector I, the modified Maxwell equations for a polar electromagnetic perturbation can be obtained by taking the $\omega=0$ limit of \cref{sec1max1,sec1max2,sec1max3}, \ie\
\begin{subequations}\be
&f u_1'' + f \left(\phi'-\frac{2 \ell^2 \mathcal{L}_{FF}}{r^5 \mathcal{L}_F}\right) u_1' + \left(\frac{2 \ell^2 f \mathcal{L}_{FF}}{r^6 \mathcal{L}_F}-\frac{r f \phi'+l(l+1)}{r^2}\right) u_1 - \frac{l (l+1) \ell}{r^3}\,h_0 =0\,,\label{sec1max1.static}\\
& u_2 + \frac{\ell f}{r}\,h_1 = 0\,,\label{sec1max2.static}\\
& r u_2' - \left(\frac{2 \ell^2 \mathcal{L}_{FF}}{r^4 \mathcal{L}_F} + r \phi' + 1\right) u_2 + \ell  f h_1' + \ell \left[f' - f \left(\frac{2 \ell^2 \mathcal{L}_{FF}}{r^5 \mathcal{L}_F}+\phi'+\frac{2}{r}\right)\right] h_1 = 0\,,\label{sec1max3.static}
\ee\end{subequations}
while the Einstein equations for an axial gravitational perturbation can be obtained by taking the $\omega=0$ limit of \cref{AxialEq13,AxialEq23,AxialEq34}, \ie\
\begin{subequations}\be
&f h_0'' + f \phi' h_0' -\frac{2 r^2 f \left(r \phi'+1\right)+(l-1)(l+2) r^2+4 \ell^2 \mathcal{L}_F}{r^4}\,h_0  - \frac{4 \ell \mathcal{L}_F}{r^3}\,u_1 = 0\,,\label{AxialEq13.static}\\
& \frac{(l-1)(l+2) r^2+4 \ell^2 \mathcal{L}_F}{r}\,h_1  + \frac{4 \ell \mathcal{L}_F}{f}\,u_2 = 0\,,\label{AxialEq23.static}\\
&f h_1' + \left(f'-f \phi'\right)h_1 = 0\,.\label{AxialEq34.static}
\ee\end{subequations}

\Cref{sec1max2.static} can be solved for $u_2$, making \cref{sec1max3.static} automatically satisfied.
However, this solution is incompatible with \cref{AxialEq23.static,AxialEq34.static} unless $u_2=h_1=0$.
The two equations that govern static perturbations in sector I are therefore \cref{sec1max1.static,AxialEq13.static}.

\medskip
In sector II, the static Klein--Gordon and modified Maxwell equations can be obtained by the $\omega=0$ limit of \cref{KGsect2_a,Maxwellsector2},
\be
&\delta\Phi''
-\left(\phi'-\frac{f'}{f}\right)\delta\Phi'
+\left(\frac{\phi'}{r} - \frac{l(l+1) + r f' - r^2 V_{\Phi\Phi}}{r^2 f}\right)\delta\Phi
+\frac{r V_\Phi H_2}{f}
+\frac{r(H_0'-H_2'+2K')}{2}\,\Phi' =0\,,\label{KGsect2_a.static}\\
&u_4''
+\left(\frac{f'}{f}-\frac{2 \ell^2 \mathcal{L}_{FF}}{r^5
\mathcal{L}_F}-\phi'\right) u_4'
- \frac{l (l+1)}{r^2 f}\left(1  + \frac{\ell^2 \mathcal{L}_{FF}}{r^4 \mathcal{L}_F}\right)u_4
-\frac{\ell (H_0+H_2)}{2 r^2 f}
+\frac{\ell}{r^2 f}\left(1+\frac{\ell^2 \mathcal{L}_{FF}}{r^4 \mathcal{L}_F}\right)K = 0
\,.\label{Maxwellsector2.static}
\ee

Again, the $\theta\varphi$ component of the perturbed gravitational equation requires $H_2=-H_0$, but the $tr$ component further imposes $H_1=0$, leaving four independent gravitational equations,
\begin{subequations}\be
&f K''
+\left(\frac{f'}{2}+\frac{3 f}{r}\right) K'
-\left(\frac{(l-1)(l+2)}{2 r^2}+\frac{2\ell^2\mathcal{L}_F}{r^4}\right)K
+\frac{f H_0'}{r}
+\frac{2 r f'+2f\left(1-r\phi'\right)+l(l+1)}{2 r^2}\,H_0\nonumber\\
&\qquad+\frac{2 l (l+1)\ell \mathcal{L}_F}{r^4}\,u_4
+\frac{4 f\sqrt{\pi r\phi'}}{r^2}\,\delta\Phi'
+\frac{8\pi r^2 V_\Phi - 4 f\sqrt{\pi r\phi'}}{r^3}\,\delta\Phi = 0\,,\label{PolarGrav11.static}\\
&\left(\frac{f'}{2}+\frac{f}{r}-f\phi'\right)K'
-\left(\frac{(l-1)(l+2)}{2 r^2}+\frac{2\ell^2\mathcal{L}_F}{r^4}\right)K
+\frac{f H_0'}{r}
-\frac{2 f\left(r\phi'-1\right) - 2 r f' +l(l+1)}{2 r^2}\,H_0
\nonumber\\
&\qquad+\frac{2 l (l+1)\ell \mathcal{L}_F}{r^4}\,u_4
-\frac{4 f\sqrt{\pi r\phi'}}{r^2}\,\delta\Phi'
+\frac{8\pi r^2 V_\Phi + 4 f\sqrt{\pi r\phi'}}{r^3}\,\delta\Phi = 0\,,\label{PolarGrav22.static}\\
&H_0'
+\left(\frac{f'}{f}-2\phi'\right)H_0
+K'
-\frac{4\ell \mathcal{L}_F}{r^2}\,u_4'
+\frac{8\sqrt{\pi r\phi'}}{r^2}\,\delta\Phi = 0\,,\label{PolarGrav24.static}\\
&f H_0''
+\left(2 f'+\frac{2f}{r}-3f\phi'\right)H_0'
+\left(\frac{2\left(r f'-1\right)-2 f\left(r\phi'-1\right)}{r^2}+\frac{4\ell^2\mathcal{L}_F}{r^4}\right)H_0
+f K''
+\left(f'+\frac{2f}{r}-f\phi'\right)K'\nonumber\\
&\qquad+\left(\frac{4\ell^2\mathcal{L}_F}{r^4} + \frac{4\ell^4\mathcal{L}_{FF}}{r^8}\right)K
-4 l (l+1)\ell\left(\frac{\mathcal{L}_F}{r^4}+\frac{\ell^2\mathcal{L}_{FF}}{r^8}\right)u_4
+\frac{8 f\sqrt{\pi r\phi'}}{r^2}\,\delta\Phi'
+\frac{16\pi r^2 V_\Phi - 8 f\sqrt{\pi r\phi'}}{r^3}\,\delta\Phi
= 0\,.\label{PolarGrav33.static}
\ee\end{subequations}

We now solve \cref{PolarGrav22.static,PolarGrav24.static} for $K$ and $K'$.
The derivative of \cref{PolarGrav24.static} with respect to $r$, together with the background equations and \cref{Maxwellsector2.static}, gives an equation for $K''$.
Using these relations and the background equations, \cref{PolarGrav33.static} is identically satisfied.
The gravitational perturbation are then described by a non-homogenous differential equation for $H_0$,
\be
f H_0'' - \eta_1 H_0' - \eta_2 H_0 + J_3 = 0\,,
\ee
where
\be
\eta_1 &= \frac{4\ell^2}{\Delta}\left(f'-2 f\phi'\right)\left(\mathcal{L}_F+\frac{\ell^2\mathcal{L}_{FF}}{r^4}\right)-f'+f\left(\phi'-\frac{2}{r}\right),\\
\eta_2 &= \frac{4\ell^2}{\Delta}\left(\mathcal{L}_F+\frac{\ell^2\mathcal{L}_{FF}}{r^4}\right)\left(\frac{l (l+1)}{r^2}-4 f'\phi'+\frac{f'^2}{f}+\frac{2 f\left(r\phi'-1\right)\left(2 r\phi'+1\right)}{r^2}\right) - 2 f'\left(2\phi'+\frac{1}{r}\right)\nonumber\\
&\phantom{=}+\frac{f'^2}{f}+\frac{2 f\left(r\phi'+1\right)\left(2 r\phi'-1\right)}{r^2}+\frac{l^2+l+2}{r^2}-\frac{4\ell^2\mathcal{L}_F}{r^4}\,,\\
J_3 &= -\frac{4\ell}{r^2}\left[\frac{2 f\mathcal{L}_F}{r}-\left(f'-\frac{2 f\left(r\phi'-1\right)}{r}\right)\left(\frac{4\ell^4\mathcal{L}_F\mathcal{L}_{FF}}{\Delta r^4}+\frac{4\ell^2\mathcal{L}_F^2}{\Delta}+\mathcal{L}_F\right)\right] u_4'
-\frac{4 \ell (l-1)l (l+1)(l+2)}{r^2 \Delta}\left(\mathcal{L}_F +\frac{\ell^2\mathcal{L}_{FF}}{r^4}\right)u_4\nonumber\\
&\phantom{=} -\frac{32\ell^2 f\sqrt{\pi r\phi'}}{r^2\Delta}\left(\mathcal{L}_F+\frac{\ell^2\mathcal{L}_{FF}}{r^4}\right)\delta\Phi'\nonumber\\
&-\frac{4}{r}\left[\frac{8\ell^2}{\Delta}\left(\mathcal{L}_F+\frac{\ell^2\mathcal{L}_{FF}}{r^4}\right)\left(\sqrt{\pi r\phi'}\left(\frac{f'}{r}+\frac{f\left(1-2 r\phi'\right)}{r^2}\right)-2\pi V_\Phi\right)-\frac{\sqrt{\pi} f\left[r\phi''+\phi'\left(2 r\phi'+3\right)\right]}{r\sqrt{r\phi'}}
\right]\delta\Phi\,,\\
\Delta &= (l-1)(l+2)r^2 + 4\ell^2\mathcal{L}_F\,.
\ee

Once the solutions for $H_0$, $u_4$ and $\delta\Phi$ are known, the other metric function perturbation is given by
\be
K = -\frac{r^2 f\left[l(l+1) - 4 r^2 f'\phi'+2 f\left(r\phi'-1\right)\left(2 r\phi'+1\right)\right] + r^4 f'^2}{\Delta f}\,H_0
-\frac{r^4 \left(f'-2 f\phi'\right)}{\Delta}\,H_0'
+\frac{4\ell l (l+1) \mathcal{L}_F}{\Delta}\,u_4\nonumber\\
+\frac{4 \ell r \mathcal{L}_F \left[r f'+2f(1-r\phi')\right]}{\Delta}\,u_4'
+\frac{16\pi r^3 V_\Phi + 8 r\sqrt{\pi r\phi'}\left[f\left(2 r\phi'-1\right)-r f'\right]}{\Delta}\,\delta\Phi
-\frac{8 r^2 f\sqrt{\pi r\phi'}}{\Delta}\,\delta\Phi'\,.
\ee

\bibliography{QNMs}
\end{document}